\documentclass[useAMS,usenatbib]{mn2e} 
\usepackage{epsfig}
\usepackage{psfrag,graphicx,amsmath,multicol}
\usepackage[dvipsnames]{xcolor}
\definecolor{webgreen}{rgb}{0,.5,0}
\definecolor{webbrown}{rgb}{.6,0,0}
\usepackage[pdftex,pdfpagelabels]{hyperref}
\hypersetup{%
   colorlinks=true,hyperfootnotes=false,%
   breaklinks=true,%
   plainpages=false, bookmarksnumbered, bookmarksopen=true,
bookmarksopenlevel=1,%
   urlcolor=webbrown, linkcolor=RoyalBlue, citecolor=webgreen,%
   pdftitle={PUT TITLE HERE},%
   pdfauthor={\textcopyright\ authors},%
   pdfsubject={},%
   pdfkeywords={},%
   pdfcreator={pdfLaTeX},%
   pdfproducer={LaTeX with hyperref}%
}

\def\bm#1{\mbox{\boldmath $#1$}}
\newcommand{\comments}[1]{}
\newcommand       \be           {\begin{equation}}
\newcommand       \ee           {\end{equation}}
\newcommand       \ba           {\begin{eqnarray}}
\newcommand       \ea           {\end{eqnarray}}
\newcommand       \grad         {\nabla}
\newcommand   \lan          {\langle}
\newcommand       \ran          {\rangle}

\newcommand       \apj          {ApJ}
\newcommand       \apjl         {ApJL}
\newcommand       \apjs         {ApJS}

\newcommand       \nat          {Nature}
\newcommand       \mnras        {MNRAS}
\newcommand       \araa         {Ann. Rev. Astr. Astr.}

\def\msun{\rm \ M_\odot}
\def\mpy{\rm \ M_\odot~{\rm yr^{-1}}}

\def\kent{\rm \ keV~cm^{2}}
\def\lesssim{\mathrel{\hbox{\rlap{\hbox{\lower4pt\hbox{$\sim$}}}\hbox{$<$}}}}
\def\gtrsim{\mathrel{\hbox{\rlap{\hbox{\lower4pt\hbox{$\sim$}}}\hbox{$>$}}}}

\defcitealias{mcc11}{Paper~I}

\title[Feedback Regulation of Hot Halos]
{Thermal Instability \& the Feedback Regulation of Hot Halos in Clusters, Groups, and Galaxies}   
\author[P.\ Sharma, M.\ McCourt, E.\ Quataert, I.~J.\ Parrish]
{Prateek Sharma\thanks{Einstein Fellow~(psharma@astro.berkeley.edu)}, Michael McCourt, Eliot Quataert, \& Ian J. Parrish\\ 
Astronomy Department and Theoretical Astrophysics Center,
601 Campbell Hall, University of California, Berkeley, CA 94720, USA}

\begin{document}

\pagerange{\pageref{firstpage}--\pageref{lastpage}} \pubyear{2011}
\maketitle
\label{firstpage}

\begin{abstract}
 We present global multi-dimensional numerical simulations of the
  plasma that pervades the dark matter halos of clusters, groups, and
  massive galaxies~(the `intracluster medium;' ICM).  Observations of
  clusters and groups imply that such halos are roughly in global
  thermal equilibrium, with heating balancing cooling when averaged
  over sufficiently long time- and length-scales; the ICM is, however,
  very likely to be {\it locally} thermally unstable.
  Using simple observationally-motivated heating prescriptions, we
  show that local thermal instability~(TI) can produce a multi-phase
  medium---with $\sim 10^4$~K cold filaments condensing out of the
  hot ICM---only when the ratio of the TI timescale in the hot plasma
 ~($t_{\rm TI}$) to the free-fall timescale~($t_{\rm ff}$) satisfies
  $t_{\rm TI}/t_{\rm ff} \lesssim 10$. This criterion {\em
    quantitatively} explains why cold gas and star formation are
  preferentially observed in low-entropy clusters and groups.  In
  addition, the interplay among heating, cooling, and TI reduces the
  net cooling rate and the mass accretion rate at small radii by
  factors of $\sim 100$ relative to cooling-flow models. This dramatic reduction 
  is in line with
  observations. The feedback efficiency required to prevent a
  cooling-flow is $\sim 10^{-3}$ for clusters and decreases for lower
  mass halos; supernova heating may be energetically sufficient to
  balance cooling in galactic halos.  We further argue
  that the ICM self-adjusts so that $t_{\rm TI}/t_{\rm ff} \gtrsim 10$
  at all radii. When this criterion is not satisfied, cold filaments condense out
  of the hot phase and reduce the density of the ICM. 
  These cold filaments can power the black hole
  and/or stellar feedback required for global thermal balance, which
  drives $t_{\rm TI}/t_{\rm ff} \gtrsim 10$.  In comparison to
  clusters, groups have central cores with lower
  densities and larger radii. This can account for the deviations from
  self-similarity in the X-ray luminosity-temperature~($L_X-T_X$)
  relation.  The high-velocity clouds observed in the Galactic halo
  can be due to local TI producing multi-phase gas close to the virial
  radius if the density of the hot plasma in the Galactic halo is
  $\gtrsim 10^{-5}$~cm$^{-3}$ at large radii.
  \end{abstract}


\section{Introduction}

The cooling of baryons in dark matter halos is a critical driver of
galaxy formation on large scales
\citep[e.g.,][]{hoy53,ree77,whi78}. If the cooling time at the virial
radius is shorter than the free-fall time, most of the gas does not
virialize at large radii and instead remains cold and flows into the
halo along cosmological filaments \citep{bir03,ker05}. For halo masses
larger than $\sim 10^{12} \msun$~(roughly the halo mass of the Milky
Way; \citealt{xue08}), however, most of the baryons undergo a virial
shock and produce a pressure-supported intracluster medium~(ICM; for
simplicity, we refer to the virialized plasma at all halo masses as
the ICM).

Even when cooling is negligible near the virial radius, it can become
important at smaller radii where the gas density is higher.  Indeed,
the cooling time of the hot plasma at the centers of groups and
clusters is often observed to be much shorter than the age of the
system~(e.g., \citealt{cav08}). In the absence of heating, the ICM
would lose pressure support on a cooling timescale, leading to a
massive cooling-flow \citep[e.g.,][]{fab94}. However, the absence of
lower energy X-ray lines \citep[e.g.,][]{tam01,mol01,pet03} and the
lack of prodigious star formation in cluster cores
\citep[e.g.,][]{ode08,raf08} suggest that cooling is approximately
balanced by some form of heating, at least when averaged over a
cooling timescale.  Heating is also required to explain the high mass
cutoff in the galaxy mass function relative to the dark matter halo
mass function~(e.g., \citealt{cro06,bow06}).

Energetically, there are a number of heating sources that can in
principle balance cooling in hot halos: e.g., thermal conduction in
clusters \citep[e.g.,][]{zak03}, the kinetic energy of galaxies and
mini-halos \citep[e.g.,][]{kim05,dek08}, and the energy produced by a
central active galactic nucleus~(AGN) in the form of jets and bubbles
of relativistic plasma \citep[e.g.,][]{bir04}. However, most
non-feedback mechanisms are {\em globally} thermally unstable
\citep[e.g.,][]{kim03,par09}. That is, even if heating balances
cooling at a given time, small perturbations about that equilibrium
will cause run-away heating/cooling of the core.  AGN feedback,
however, can be {\em globally} thermally stable if the efficiency of
feedback is sufficiently large \citep[e.g.,][]{guo08}.  Physically,
this is because the accretion rate onto the central black hole
increases as the cooling rate in the halo increases, providing the
necessary feedback loop.  It is plausible that supernova heating can
provide a similar source of globally stable heating in galactic halos 
with ongoing star formation.

Although AGN feedback can plausibly ensure {\em global} thermal
stability of cluster cores, the ICM is very likely to be {\em locally}
thermally unstable because of the strong density dependence of the
cooling rate~(\citealt{fie65}). The situation is analogous to the hot
phase of the interstellar medium~(ISM) which is thermally unstable but
is maintained by supernova heating \citep[][]{mck77}.  While the thermal instability~(TI)
has been invoked to explain a number of observations of cool-core
clusters in the past, previous treatments have typically focused on
models without heating~(e.g., \citealt{nul86,kau09}).  Other studies,
very similar in spirit to ours, have emphasized the importance of AGN
feedback due to accretion of cold gas, but have focused on the
survival of dense, cool blobs once they form, rather than their origin
(e.g., \citealt{piz05}). Here we focus on how local TI produces cold filaments
condensing out of the hot phase for virialized halos in rough thermal balance.

Observationally there is a sharp threshold in central entropy/cooling
time below which cold gas is observed in groups and clusters~(e.g.,
\citealt{cav08,hec89,hu85,mcd10}). These cool-core clusters also preferentially show
evidence for AGN feedback and star formation~(e.g., \citealt{raf08,hic10}).
The strong correlation between cold gas indicators and the properties
of the hot ICM motivates a model in which TI is the source of much of
the cold gas in groups and clusters. The condensation of cold gas in hot
halos has also been invoked~(e.g., \citealt{mal04,som06,pee08}) as a
source of the high velocity clouds~(HVCs) of neutral hydrogen observed
in the Galactic halo~(e.g., \citealt{wak97}) and in the halo of M31
(\citealt{thi04}).

In a previous paper~(\citealt{mcc11}; hereafter \citetalias{mcc11}) we
studied the physics of local TI in gravitationally stratified plasmas,
including observationally-motivated heating prescriptions, anisotropic
thermal conduction, and magnetic fields.  We used local Cartesian
numerical simulations to isolate the key physics.  The unusual feature
of the TI in the ICM~(relative to, say, the molecular/atomic phase of the ISM) is that the growth
time of the TI in the hot plasma~($t_{\rm TI}$) can be comparable to,
or longer than, the free-fall time in the cluster potential~($t_{\rm
  ff}$).  In \citetalias{mcc11}, we showed that the ratio of these two
key timescales determines whether cold gas can condense out of the hot
ICM.  If the cooling time is short~($t_{\rm TI} \lesssim t_{\rm ff}$),
initially small density perturbations become nonlinear and cool blobs
reach the thermally stable state at $\sim 10^4$~K. These dense
gas-rich filaments then fall through the hot ICM at roughly the
free-fall rate~(as in \citealt{piz05}); they may cool further,
forming stars, and/or powering a central AGN. On the other hand, if
the TI growth time is longer than the free-fall time~($t_{\rm TI}
\gtrsim t_{\rm ff}$), the TI saturates before the perturbations become
nonlinear and the TI cannot produce a multi-phase medium starting from
small initial perturbations.  Note that this criterion for TI to
produce a multi-phase medium is analogous to the criterion for
cosmological accretion to proceed via cold streams rather than
virialized plasma.  However, our TI criterion applies at {\em all radii} in
hot halos, not just at the virial shock.

In this paper, we extend the calculations of \citetalias{mcc11} to
more realistic~(but still non-cosmological) global models of the ICM
in dark matter halos.  The most uncertain input into these
calculations is our treatment of the heating of the ICM.  Observations
of groups and clusters imply that there is an approximate balance
between heating and cooling in cluster cores.  Motivated by this, we
explore two different heating models; both
assume that the heating is isotropic, thus sidestepping the very
important but difficult problem of how energy is redistributed
throughout the ICM.  Our first model determines the heating rate from
the average cooling rate at each radius in the halo~(see also
\citetalias{mcc11}). We refer to these simulations as our `idealized
heating' simulations.  Our second model assumes that the heating of the ICM 
is proportional to the mass accretion rate at small radii.  We
refer to these simulations as our `feedback' heating simulations
(though they are also very idealized!).  Fortunately, these two models
give similar results.  This implies that there are robust properties
of hot halos that arise primarily from the interplay between local TI and
global thermal equilibrium, and which are not very sensitive to the
details of how global thermal equilibrium is maintained.

Our calculations do not include the physics necessary to model the
detailed properties of the gas-rich filaments produced by TI. In
reality magnetic fields, cosmic rays, and thermal conduction along
magnetic field lines strongly influence the geometry and energy
balance in filaments~(e.g., \citealt{fer09,sha10}).  Star formation
needs to be incorporated as an excitation mechanism for the observed
H$\alpha$ emission in at least some filaments~(e.g.,
\citealt{mcd11b}). Moreover, the physics of turbulent heating/mixing
and non-radiative cooling are probably required to explain the absence
of some of the lower energy X-ray lines~(\citealt{wer10,san10}); this
physics is not well-resolved in our simulations. A detailed study of
these processes is important, but beyond the scope of the present
paper.  Nonetheless, the simulations presented in this paper {\it do}
accurately describe the conditions under which gas-rich filaments form
out of the hot ICM; they also provide insight into the role of TI in
determining the observed density and entropy profiles of hot halos,
and in maintaining such halos in global thermal equilibrium.

In \citetalias{mcc11} we showed that magnetic fields and anisotropic
thermal conduction do not significantly alter the evolution of the TI for halos 
in thermal balance.  More specifically, conduction changes the
morphology of the resulting cold gas~(turning blobs into filaments
along magnetic field lines), but not the amount of cold gas produced
by TI.  As a result, we do not include magnetic fields and conduction
in this paper.

The interplay between local TI and gravity in a globally stable
cluster cannot be captured in a one-dimensional model because the
physics of cool overdense blobs falling in a hot pressure-supported
atmosphere is absent in 1-D.  Moreover, maintaining the pressure
support of the hot ICM is critical to preventing a massive
cooling-flow.  The TI also evolves very differently in the presence of
heating than it does in cooling-flows~(\citealt{bal89}; see also Appendix \ref{app} 
where we discuss the formation of multi-phase structure in cooling-flows in detail).  These
considerations imply that numerical studies of hot halos should use multidimensional simulations 
and should include heating.  Such simulations are the
focus of this paper.

The remainder of this paper is organized as follows. We describe the
properties of our multi-dimensional simulations in section 2. Section
3 presents the results of our `idealized' heating simulations.
Simulations with simplified models of `feedback' heating are presented
in section 4. We discuss important astrophysical implications of our work in section 5.
We conclude with a detailed summary of our paper in section 6. Readers not interested in 
simulation details may skip to section 5.

\section{Simulation Setup}

In this section we detail the equations that we solve numerically, and the initial and boundary conditions. We also describe the cooling and heating prescriptions that we use. 

\subsection{Governing Equations}

We solve the standard hydrodynamic Euler equations with heating, cooling, and external gravity:
\begin{equation}
\frac{d \rho}{d t} = - \rho \grad \cdot \bm{v},
\label{eq:cont}
\end{equation}
\begin{equation}
\rho \frac{d\bm{v}}{dt} = - \grad p  + \rho {\bm g},
\label{eq:momentum}
\end{equation}
\begin{equation}
e \frac{d}{dt} \ln~(p/\rho^\gamma) = -q^-(n,T) + q^{+}(r,t),
\label{eq:energy}
\end{equation}
where $d/dt \equiv \partial /\partial t + {\bm v} \cdot \grad$ is the Lagrangian derivative, $\rho$ is the mass density, $\bm{v}$ is the fluid velocity, $p$ is the thermal pressure, $e=p/(\gamma-1)$~(we use $\gamma=5/3$, valid for an ideal non-relativistic gas) is the internal energy per unit volume, $\bm{g}$ is the acceleration due to gravity, $q^+$ is the heating rate per unit volume, $q^-\equiv n_e n_i \Lambda(T)$ is the cooling rate per unit volume, $n_e$~($n_i$) is the electron~(ion) number density, and $\Lambda(T)$ is the cooling function.  We use a third of the solar metallicity, corresponding to a mean molecular weight per particle $\mu \equiv \rho/(nm_p)=0.62$ and a mean molecular weight per electron $\mu_e \equiv \rho/(n_em_p)=1.18$, where $n=n_e+n_i$ is the total number density and $m_p$ is the proton mass.

We model the ICM in spherical~($r,\theta,\phi$) geometry using the ZEUS-MP code \citep[][]{hay06}. We model the dark matter halo as a fixed gravitational potential with an NFW profile:
\be \Phi = -\frac{2 G M_0}{r_s} \frac{\ln~(1+r/r_s)}{r/r_s}, \ee 
where $M_0$ is the characteristic dark matter mass and $r_s$ is the scale radius \citep{nav97}. The acceleration due to gravity 
$$
{\bm g} \equiv - \frac{d\Phi}{dr} \bm{\hat{r}}
$$ 
is directed toward the center of the dark matter halo. For simplicity we do not include the contribution of the ICM or the galaxies to the gravitational potential. For our fiducial cluster model C10, we take $M_0=3.8 \times 10^{14} \msun$ and $r_s=390$~kpc. We scale the radii according to the self-similar scaling for different halo masses; $\{r,~r_s\} \propto M_0^{1/3}$. The concentration parameter $c \equiv r_{200}/r_s$~(where $r_{200}$ is the virial radius within which the mean dark matter density is 200 times the critical density of the universe) is fixed to be $c=3.3$ for all halo masses. With this choice the virial mass is $M_{200} = 1.38 M_0$.

The equations are solved using operator~(and directional) splitting into source and transport steps~(\citealt{sto92}). We combine the heating and cooling rates at each grid cell and subcycle the internal energy equation. Cooling for simulations without heating~(i.e., for cooling-flows with $q^+=0$) is not subcycled but is treated semi-implicitly~(see Eq. 7 in \citetalias{mcc11}); an explicit treatment would lead to computationally prohibitive subcycling at almost all timesteps because of short cooling times. This less accurate treatment of cooling does not affect the results much. All of the simulations are run for 5~Gyr, roughly the time between major mergers of halos at the current epoch.

\subsubsection{Initial and Boundary Conditions}

We initialize the ICM in hydrostatic equilibrium with an entropy profile\footnote{Although an entropy core provides a good fit for most clusters, some non-cool-core clusters have a high-density/low-temperature `corona' around the central BCG at $\lesssim 5$~kpc~(see \citealt{sun09}). We do not address these systems in this paper.}
\be
\label{eq:ini_ent}
K(r) \equiv \frac{T_{\rm keV}}{n_e^{2/3}}=K_0+K_{100} \left (\frac{r}{r_{100}} \right )^\alpha,
\ee
where $r_{100}=100$~kpc for our fiducial run~(C10) and scales with $M_0^{1/3}$ for other halo masses \citep[][]{cav09}. The entropy profiles that we use cover the range observed in clusters. We seed initial density perturbations such that the maximum $\delta \rho/\langle \rho \rangle \approx 0.3$, where $\delta \rho = \rho - \langle \rho \rangle$ and $\langle \rho \rangle$
is the angle-averaged density as a function of $r$. The density fluctuations~($\delta \rho/\lan \rho \ran$) are isotropic and homogeneous~(in the three-dimensional Cartesian sense), and have a one-dimensional power spectrum $\propto k^{3/2}$ 
for $4 \lesssim k r_{\rm out}/2\pi \lesssim 10$, where $r_{\rm out}$ is the outer radius~(200~kpc for the fiducial run). Though somewhat arbitrary, this choice for the initial perturbations is fairly general and enables us to initialize all of our simulations identically. For simulations with heating, the resulting quasi-steady state does not depend strongly on the initial density perturbations. By contrast, the properties of  the multi-phase gas in cooling-flows do depend on the initial density perturbations~(see Appendix \ref{app}---this is because TI does not grow exponentially in cooling-flows; \citealt{bal89}).

We perform our simulations in spherical~($r,\theta,\phi$) coordinates, where $1~{\rm kpc} \leq r \leq 200~{\rm kpc}$ for the fiducial run C10, $0 \leq \theta \leq \pi$, and $0\leq\phi \leq2 \pi$. Again, we adjust the radial range for different halo masses such that $r \propto M_0^{1/3}$.  We use a logarithmic grid in the radial direction and a uniform grid in other
coordinate directions. We fix the outer electron number density to be $n_{e, {\rm out}} = 0.0015$~cm$^{-3}$ for all halo masses, a 
value observed for Abell 2199 at 200~kpc~(see \citealt{joh02}; see also \citealt{sha09} for details of the setup). The outer radius is close to the scale radius of Abell 2199, such that the initial temperature increases monotonically with radius. We apply inflow~(outflow) boundary conditions at the outer~(inner) radial boundary, such that a steady inflow may develop; mass is not allowed to flow out of~(in to) the computational domain at the outer~(inner) boundary. Since the cooling time at the outer boundary is much longer than 5~Gyr, we do not expect our results to change significantly with a different outer boundary condition.\footnote{The outer boundary conditions do not affect the results for our simulations with heating, in which a hot pressure-supported ICM is maintained. However, the cooling-flow simulations without heating do depend on the outer boundary conditions, as we discuss in Appendix \ref{app}.} We apply reflective boundary conditions in the $\theta$ direction and periodic boundary conditions in the $\phi$ direction.

\subsection{Cooling \& Heating}

We use the cooling function $\Lambda(T)$ given by Eq.~(12) in \citet{sha10}
(corresponding to the solid line in their Fig.~1), which is a fit to the
equilibrium cooling curve of \citet{sut93} for 0.3 solar metallicity~(see
their Fig.~8).  Hydrogen starts to recombine below $\sim 10^4$\,K; this
abruptly reduces the cooling rate and creates a thermally stable ``cold''
phase.  The ICM is locally thermally unstable at all temperatures above $10^4$\,K, but
we show later~(Fig.~\ref{fig:lum_vs_T}) that the combined effect of cooling,
heating, and gravity results in a strongly bimodal temperature distribution in
the plasma.  We therefore speak of distinct ``hot'' and ``cold'' phases.

While the cooling function for the ICM is fairly well-known, the heating of
the ICM is not yet understood.  For concreteness we choose a heating rate
such that the energy deposited per unit volume is constant~($q^+ \propto
\rho^0$).  The simulations in \citetalias{mcc11} demonstrate that the results
are not sensitive to this choice, as long as time is normalized to the thermal
instability timescale.  Note that, while the cooling rate~($q^-\equiv
n_en_i\Lambda$) is a thermodynamic state function, the heating rate $q^+(r,t)$
explicitly depends on radius and time.  Our idealized heating prescription and
our feedback heating prescription differ in the form of $q^+(r,t)$;
i.e., the two models represent different assumptions about how the feedback
energy is distributed in the ICM.  In our
idealized simulations, we distribute the heating to instantaneously balance
cooling at all radii in the plasma and in our feedback simulations we make the
heating proportional to the mass accretion rate at small radii.

\subsubsection{Idealized Heating}

\begin{figure*}
\centering
\includegraphics[scale=0.31]{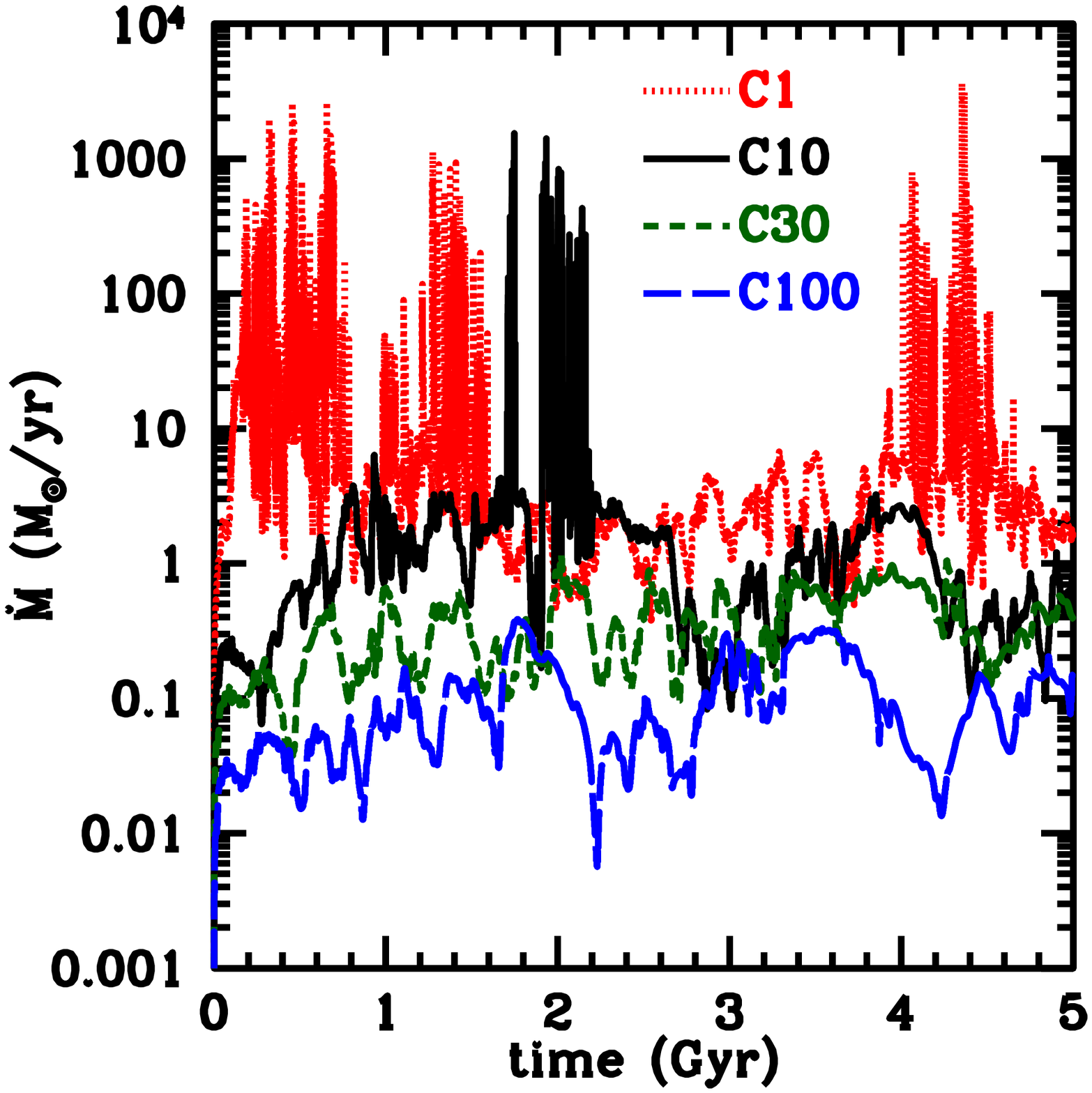}  \includegraphics[scale=0.31]{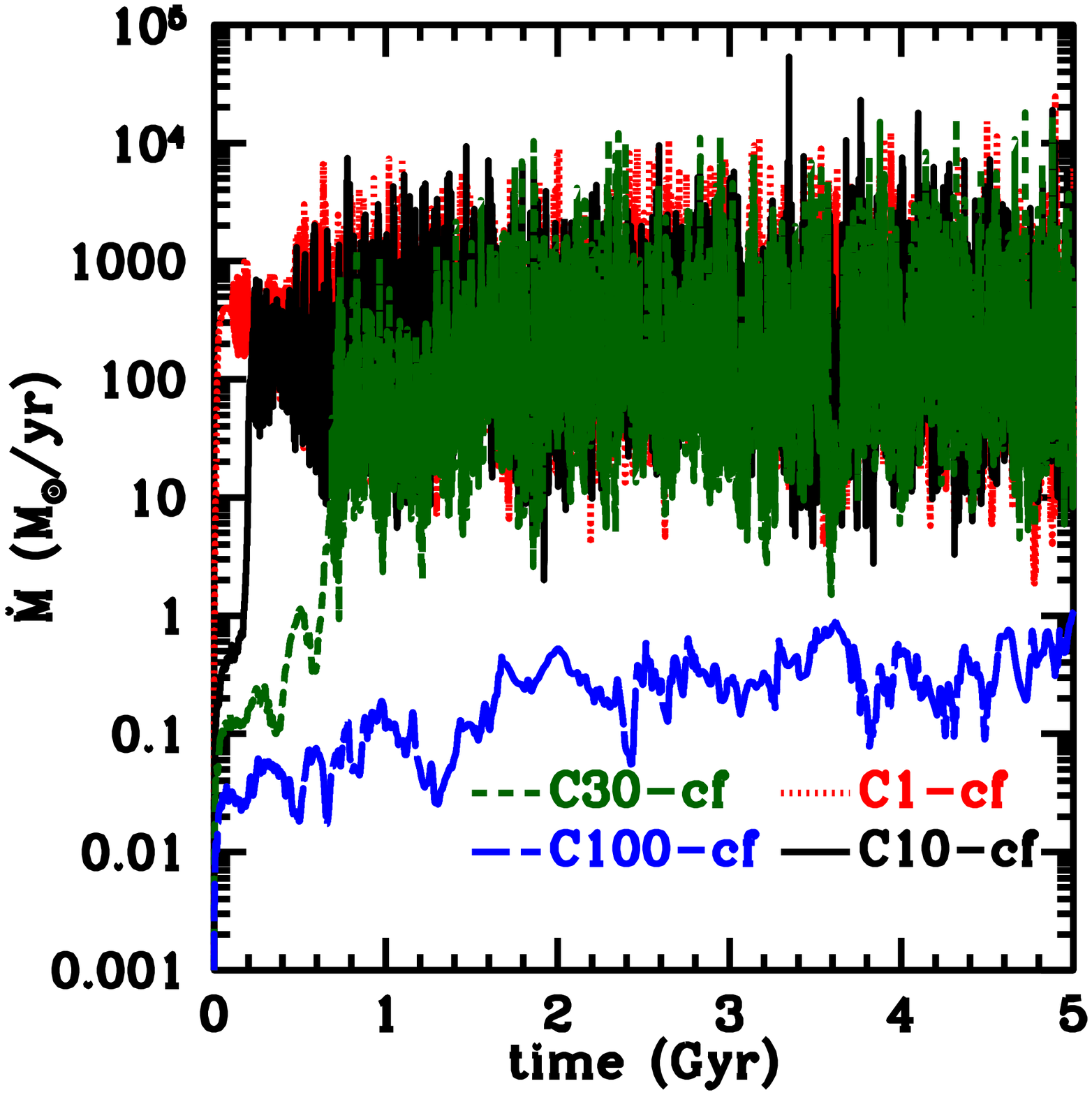}
\caption{Mass accretion rate~(in $\mpy$) through the inner boundary~(at 1~kpc) for different initial entropy profiles as a function of time for our cluster runs with idealized heating~({\em left} panel) and without heating~(i.e., cooling-flows; {\em right} panel). In the simulation labels the number following `C' is the initial core entropy~($K_0$ in $\kent$). For simulations with heating the mass accretion rate at late times is reduced because the cold filaments are accreted and a hot ICM with $t_{\rm TI}/t_{\rm ff} \gtrsim 10$ is maintained by heating. All of the low entropy cooling-flow runs look similar at late times and reach a cooling catastrophe. The cooling time for runs with $K_0 \gtrsim 100 \kent$ is long enough  that a catastrophic cooling-flow does not occur within 5~Gyr even without heating. 
\label{fig:mdot}}
\end{figure*}

For simulations with idealized heating~(Tables \ref{tab:tab1} \& \ref{tab:tab2}) the heating rate at any point in the plasma~(in erg~cm$^{-3}$~s$^{-1}$) is equal to to the angle-averaged cooling at that radius:
\be
\label{eq:heat}
q^+(r,t) = \frac{1}{4\pi} \int_{\theta=0}^{\pi} \int_{\phi=0}^{2\pi} 
           q^-[n({\bm{r}},t),T({\bm{r}},t)] \sin \theta d\theta d\phi.
\ee
This imposes perfect thermal equilibrium at all radii and at all times. Although at first Eq. \ref{eq:heat} appears overly idealized, it provides a useful framework for studying the thermal instability. Furthermore, the results of \citetalias{mcc11} and section \ref{sec:robust} show that introducing even large~($\sim 100\%$ or more) fluctuations about the heating rate in Eq. \ref{eq:heat} does not significantly change the resulting properties of the hot halo~(so long as global thermal equilibrium is approximately maintained over a few cooling timescales).

\subsubsection{Feedback Heating}

In our feedback simulations we set the heating rate
\be
\label{eq:heat_fb}
q^+(r,t) = \frac{K_n}{4\pi r_{\rm cf}^3} 
           \frac{(r/r_{\rm cf})^{\alpha}}{[1 +~(r/r_{\rm cf})^\beta]}
           \epsilon_{\rm fb} \dot{M}_{\rm in} c^2,
\ee
where $K_n$ is a normalization constant such that the integral of the heating rate over the whole volume $\int q^+ 4 \pi r^2 dr = \epsilon_{\rm fb} \dot{M}_{\rm in} c^2$, $\epsilon_{\rm fb}$ is the feedback  heating efficiency, $\dot{M}_{\rm in}$ is the instantaneous mass accretion rate through the inner radius, $c$ is the speed of light, $r_{\rm cf}$ is a cutoff radius for heating, and $\alpha$ and $\beta$ are parameters determining the heating profile. We set the cutoff radius equal to the cooling radius where the cooling time is 5~Gyr; this gives $r_{\rm cf}=100$~kpc for clusters~($M_0=3.8 \times 10^{14} \msun$) and $r_{\rm cf}=70$~kpc for groups~($M_0=3.8 \times 10^{13}\msun$). The cutoff radius must be larger than the cooling radius to prevent a cooling catastrophe at late times due to cooling beyond the cutoff radius. If the cutoff radius is too much larger than the cooling radius, however, we require a higher feedback efficiency to balance cooling in the core because  the feedback deposits a lot of energy beyond the cooling radius, where it is not required. The parameters $\alpha=-1.5$ and $\beta=3.5$ are chosen such that most of the energy is deposited near $r_{\rm cf}$. The choice $\alpha = -1.5$ ensures that a relatively flat density/entropy core is obtained, as observed; for a heating profile steeper~(shallower) than this we obtain an inner density profile which increases~(decreases) with radius. The results are not very sensitive to the exact choice of the parameters $\alpha$, $\beta$, and $r_{\rm cf}$. We discuss the dependence on $\epsilon_{\rm fb}$ in detail in section \ref{sec:fb}.

Since the heating rate in both models~(Eqs. \ref{eq:heat} \& \ref{eq:heat_fb}) is constant at all angles for a fixed radius and time~($q^+ \propto \rho^0$), and since cooling is sensitively dependent on density~($\propto \rho^2 \Lambda[T]$), fluid elements cooler and denser than average will cool further; though the plasma is globally in thermal equilibrium, it is locally thermally unstable.

\subsection{Important Timescales}
In this section we define the most important timescales in our analysis. The TI growth time~(inverse of the exponential growth rate) in the isobaric regime is \citep[e.g., see Eq. 8 in][]{sha10}
\be 
\label{eq:tTI}
t_{\rm TI} =  \frac{5}{3} \frac{t_{\rm cool}}{(2 - d \ln \Lambda/d\ln T) },
\ee
where 
\be
\label{eq:tcool}
t_{\rm cool} \equiv \frac{3}{2} \frac{n k_B T}{n_en_i\Lambda}. 
\ee
In Eq. \ref{eq:tTI} we have assumed volumetric heating~($q^+ \propto \rho^0$) and we have ignored thermal conduction which suppresses the instability at small scales \citep[][]{fie65}. This approximation is justified because thermal conduction is anisotropic with respect to the magnetic field~(see \citetalias{mcc11} for more details). For free-free cooling~(with $\Lambda \propto T^{1/2}$) relevant for clusters, $t_{\rm TI}$ =~(10/9)$t_{\rm cool}$. This timescale depends on how the feedback energy is thermalized in
the ICM: if instead $q^+\propto \rho$, the TI timescale is longer, $t_{\rm TI}
=~(10/3) t_{\rm cool}$, because heating partially compensates cooling for an
overdense blob. We also define the free-fall time 
\be \label{eq:tff} t_{\rm ff} \equiv \left ( \frac{2 r}{g} \right )^{1/2}, \ee
where $g\equiv d\Phi/dr$ is the acceleration due to gravity. As we discuss later, the ratio $t_{\rm TI}/t_{\rm ff}$ plays a crucial role in the physics of the ICM---if this ratio is small cold filaments condense out of the hot phase, resulting in an abrupt rise in the accretion rate and in the associated feedback heating.

\section{Simulations with Idealized Heating}
In this section, we describe the numerical results from simulations with our
idealized heating prescription. These idealized heating runs, in which heating at each radius balances angle-averaged cooling, are motivated by the observed thermal equilibrium in cluster cores. The identification that cold filaments are due to TI in a globally stable ICM is easily motivated in this setup: the ICM evolves only because we seed the local TI with density
perturbations. Although this idealized heating function appears to be finely tuned, we find that the results are quite insensitive to large temporal and spatial perturbations in heating on top of this exact balance~(see section \ref{sec:robust} \& \citetalias{mcc11}). Moreover, the more realistic feedback simulations discussed in section \ref{sec:fb} reach a quasi-steady state similar to these idealized simulations.

In section \ref{sec:TI} we discuss the effects of our heating ansatz on the mass accretion rate and the cooling to sub-virial temperatures; we also compare our results to a cooling-flow.  In \ref{sec:fid} we discuss our fiducial run C10 in detail. Section \ref{sec:self_reg} shows that TI and heating leads to a lower limit on $t_{\rm TI}/t_{\rm ff}$ and the core entropy. In section \ref{sec:diff_mass} we show that TI affects larger radii in lower mass halos, and that there is a significant lack of plasma below $\sim 1/2-1/3$ of the virial temperature in simulations with heating. We end this section with a discussion of the robustness of our results to large variations in heating on top of our thermal balance and to changes in resolution~(section \ref{sec:robust}). Table \ref{tab:tab1} lists the two-dimensional simulations with idealized heating~(see Eq. \ref{eq:heat}). Table \ref{tab:tab2} lists simulations with temporal and spatial heating fluctuations, three-dimensional simulations, and simulations at different resolutions.

\subsection{Effect of Heating on Cooling \& Accretion}
\label{sec:TI}

To assess the effect of heating on the structure of the cooling core we also carried out a few runs without heating~($q^+=0$; i.e., cooling-flows). In this section we compare the results of simulations with idealized heating to cooling-flow simulations. The properties of the cooling-flow simulations are discussed in more detail in Appendix \ref{app}.

\subsubsection{Suppression of the Mass Accretion Rate via Heating}

Figure~\ref{fig:mdot} compares the mass accretion rate through the inner radius for cluster simulations~($M_0=3.8\times 10^{14} \msun$; prefixed with C in Table \ref{tab:tab1}) with idealized heating~({\em left} panel) and without heating ~({\em right} panel) for different initial entropy profiles. The idealized heating runs with initial core entropy $K_0 \geq 30 \kent$ have a mass accretion rate $< 1\mpy$ at all times. Runs with lower core entropies~(C10 and C1) have much larger instantaneous mass accretion rates~(up to 100s of $\mpy$), but only for short times; more typically they show accretion rates between 1 and 10 $\mpy$, in broad agreement with accretion rates inferred from X-ray spectral modeling~(see \citealt{pet06} for a review on X-ray spectroscopy of clusters) and from the star formation rates in the central galaxy \citep[e.g.,][]{ode08}. The cooling-flow runs are qualitatively different from the idealized heating simulations.  After a few cooling times the mass accretion rate for the cooling-flow runs is consistently larger than 100 $\mpy$. Only when the central cooling time is longer than 5~Gyr~(as is the case for $K_0=100 \kent$) does one escape this conclusion. Thus, as intended, our idealized heating prescription prevents a cooling catastrophe even for runs with short cooling times.

\begin{figure}
\centering
\includegraphics[scale=0.42]{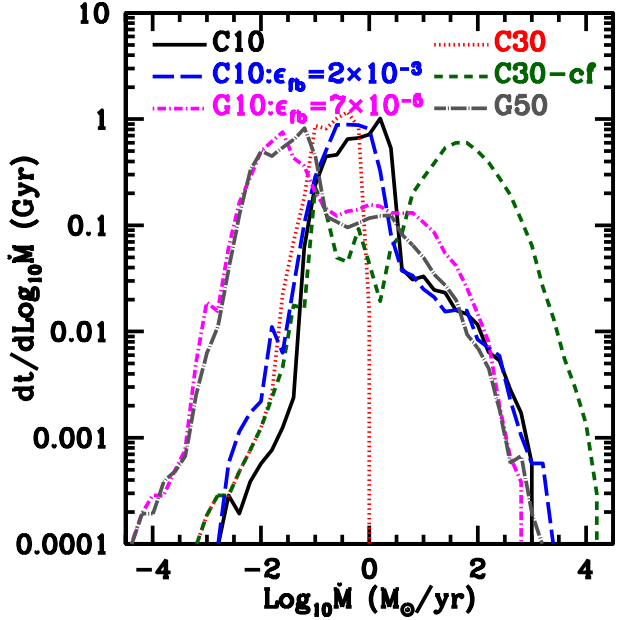}
\caption{The mass accretion rate through the inner radius~(e.g., in Fig.~\ref{fig:mdot}) shows large variations  as a function of time, especially when cold filaments cross the inner boundary. This plot represents the same information as Fig.~\ref{fig:mdot}, but shows the time interval spent accreting with a given mass accretion rate; we use 50 logarithmically spaced bins in the total mass accretion rate~($\dot{M}$) from $10^{-5}$ to $10^5\mpy$. While cooling-flows accrete at large rates~($\sim 100\mpy$) in the cold phase, simulations with heating~(even the low entropy run C1; not shown in this Figure) spend most of their time accreting in the hot phase with $\dot{M} \lesssim 1\mpy$. The results for our idealized heating model and the feedback heating simulations are very similar for both clusters and groups.  While the mass accretion rate in the hot phase is smaller for smaller mass halos, the mass accretion rate is similar whenever $\dot{M} \gtrsim {\rm few}\mpy$; accretion with $\dot{M} \gtrsim {\rm few}\mpy$ is mainly due to the infalling cold gas. 
\label{fig:time_vs_mdot}}
\end{figure}

Figure~\ref{fig:time_vs_mdot} shows an alternate view of the fundamental difference between cooling-flows and clusters in global thermal balance. Here we show the time spent at a given $\dot{M}$ for simulations with and without heating. While the mass accretion rate is $\sim 100\mpy$ at most times for cooling-flows, simulations with idealized heating primarily accrete at $\lesssim 1\mpy$.  Even for the heating runs that show $\dot{M} \gtrsim 100\mpy$ instantaneously, the time spent in such a phase is only 10~Myr.\footnote{The Eddington rate for a $10^9 \msun$ black hole is $\sim 10 \mpy$~(assuming a radiative efficiency of 0.1). Thus, if most of the mass accreted in form of cold filaments with $\dot{M} \gtrsim 1 mpy$ can reach the supermassive black hole, it can power a luminous quasar for a short duration~(e.g., see \citealt{cio97}). Since our simulations do not include star formation and accretion physics we cannot explore this possibility in detail.} This high accretion rate arises only in low entropy simulations~(see Table \ref{tab:tab1}) and represents cold filaments which condense out of the hot phase and cross the inner boundary.  Simulations with intermediate
entropies~(such as the run C30) do not develop multi-phase gas~(see Table
\ref{tab:tab1}) and never have large accretion rates; $\dot{M} \lesssim 1 \mpy$ at all times. Thus, large accretion rates in simulations with our idealized heating prescription only occurs via cold gas. We show later that this result is not sensitive to our
choice for the heating.

The absence of cold filaments in our idealized heating simulations with $K_0 \gtrsim 30 \kent$ appears surprising at first because the cooling time in these clusters~($\approx 1$~Gyr) is shorter than the run-time---since
we initialized the ICM with large density perturbations, one might expect the
TI to become strongly nonlinear and to produce multi-phase gas. The cooling time in these simulations is much longer than
the free-fall time, however, and gravity significantly alters the saturation
of the TI in this limit.The density perturbations saturate at a small amplitude~($\delta \rho/\rho < 1$) because the effective gravity felt by slightly overdense blobs~($g \delta \rho/\rho$) produces strong shear which mixes the overdense blobs with the hot phase within a cooling time. Thus, overdense fluid elements do not cool all the way to the thermally stable phase. By contrast, when the TI timescale is shorter than the free-fall time~($t_{\rm TI}/t_{\rm ff} < 1$) gravity has essentially no effect on the TI. The TI develops and saturates over a few cooling times, and only after a free-fall time do the cold filaments fall in toward the center~(see \citetalias{mcc11} for a more detailed discussion of the physics of TI in the presence of gravity). In our setup, cold filaments condense out from the hot phase, only if the ratio $t_{\rm TI}/t_{\rm ff} \lesssim 10$ locally; we discuss this more quantitatively below.

\subsubsection{Cooling to Sub-Virial Temperatures}
\label{sec:subv}
\begin{figure}
\centering
\includegraphics[scale=0.42]{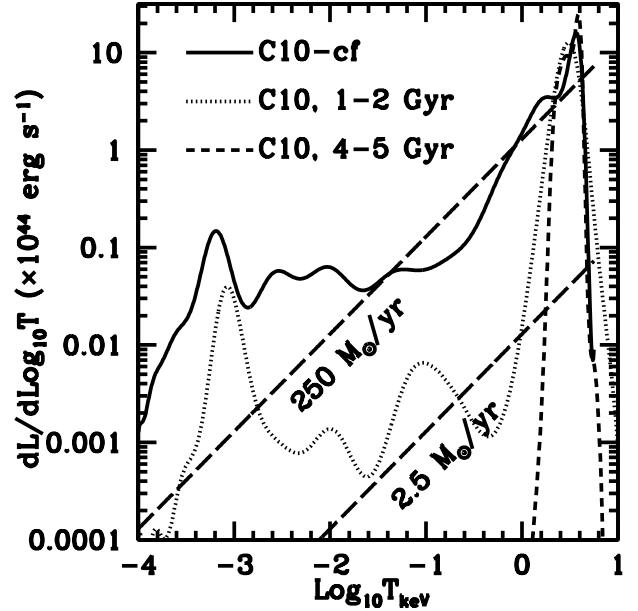}
\caption{Luminosity per logarithmic temperature bin~($dL/{\rm Log}_{10}T$; such that the total luminosity is given by the area under the curve) as a function of ${\rm Log}_{10}T_{\rm keV}$ for the cooling-flow run~(C10-cf) in the quasi-steady state, for the idealized heating run C10 averaged from 1--2~Gyr~(during which it shows cold filaments), and for C10 averaged from 4--5~Gyr~(when the filaments are absent). We use 1000 bins in ${\rm Log}_{10}T_{\rm keV}$, and the simulation data are averaged in time and fit by a high-order polynomial. The results expected from steady cooling-flows~($dL/dT = [5/2] k_B T \dot{M}/\mu m_p$; e.g., \citealt{fab94}) with $\dot{M} =250\mpy$ and 2.5 $\mpy$~(long dashed lines) are shown for comparison. 
\label{fig:lum_vs_T}}
\end{figure}

The effect of heating is also manifest in the temperature structure of the plasma. 
Figure~\ref{fig:lum_vs_T} shows the luminosity per logarithmic temperature bin for simulations with~(C10) and without~(C10-cf) heating. We
show the simulation with heating at two times, one at which there is
significant cold gas via TI~(1--2~Gyr) and one at which there is not~(4--5~Gyr). Even when filaments are present, the luminosity between $\sim 10^4$\,K and
$10^7$\,K is a factor of 10--100 times smaller for the run with heating than
for a cooling-flow. This is consistent with observations that find a striking absence of the low-energy X-ray lines expected in a cooling-flow~(e.g., \citealt{pet06}). The luminosity at $\approx 10^4$~K~(below which the plasma becomes thermally stable) is also smaller for simulation with heating, but only by a factor $\sim 3$. There is a significant lack of plasma at the intermediate temperatures~(0.001--1~keV) for simulations in thermal balance because   gas blobs at intermediate temperatures are thermally unstable and run away either to the hot or the cold phase. Figure~\ref{fig:lum_vs_T} also shows that the heating simulations do not show any gas below $\approx 1$~keV  form 4--5~Gyr~(when  $t_{\rm TI}/t_{\rm ff} \gtrsim 10$) because there is no cold gas produced by TI. 

The luminosity vs.\ temperature for our cooling-flow simulation in Figure~\ref{fig:lum_vs_T} closely matches the analytic cooling-flow model between 0.1 to 1~keV~($dL/dT = [5/2] k_B T \dot{M}/\mu m_p$; e.g., \citealt{fab94}), with a cooling rate $\dot{M}= 250\mpy$~(a value similar to the accretion rate in cooling-flow simulations; see Table \ref{tab:tab1}). Without a heat
source, the plasma cools almost monolithically; our cooling-flow simulations
thus possess roughly equal mass per cooling time interval. When heating balances cooling, however,  the ICM does not cool monolithically. 
Slightly overdense cool blobs are therefore able to mix with the hotter ambient plasma as they fall in. Thus cooling from 1 to 0.1~keV follows a random walk in cooling/heating rather than cooling monolithically~(\citealt{sok03} envisage a similar mechanism of a `moderate cooling-flow') and the slope of the luminosity function is much steeper. Moreover, the amount of gas cooling
through lower X-ray temperatures is smaller, as observed, because of
reheating.  Therefore, in simulations with heating the net cooling rate below
1~keV, which is approximately equal to the average mass accretion rate in the cold phase at small radii, is 10--100 times smaller than a cooling-flow.\footnote{The exact
suppression of luminosity relative to a cooling-flow depends on the ratio
$t_{\rm TI}/t_{\rm ff}$.  For a small $t_{\rm TI}/t_{\rm ff}$, e.g., at early
times in the low entropy run C1 the luminosity below 1~keV is closer to the
cooling-flow value.  For larger $t_{\rm TI}/t_{\rm ff}$, the luminosity below 1~keV can be vanishingly small, as shown by the C10 results at 4--5~Gyr in Figure~\ref{fig:lum_vs_T}.} However, even in the presence of heating, once gas manages to cool below $\sim 1$~keV, it runs away all the way to $10^4$~K. Thus our idealized heating run resembles a much-reduced cooling-flow below $\sim 1$~keV~(cooling is reduced by $\sim 100$; compare with the cooling-flow prediction for $\dot{M}= 2.5\mpy$).

\begin{figure}
\centering
\includegraphics[scale=0.37]{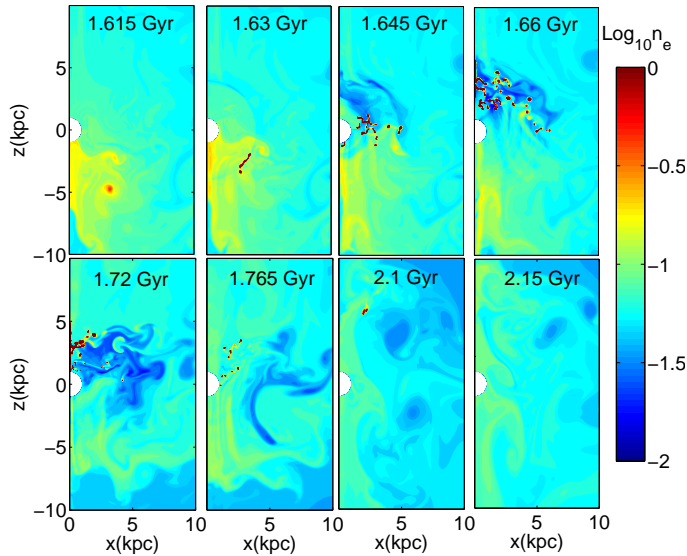}
\caption{Snapshots of the electron number density in the inner 10~kpc for the fiducial simulation~(C10) at different times around when cold filaments condense out of the hot phase. The limits on the electron density contour plots only go from $0.01$ to 1~cm$^{-3}$ but the coolest/densest plasma is extremely cold~($\sim 10^4$~K) and dense~($\sim 100$~cm$^{-3}$). Cold filaments condense out of the hot phase and eventually fall in toward the center; however, the cold gas does not just free-fall radially inwards and can even move outwards in radius for some time. \label{fig:2D}}
\end{figure}

The cooling-flow prediction using $\dot{M} = 250 \mpy$ in Figure~\ref{fig:lum_vs_T} matches our cooling only simulation for temperatures above 0.1~keV. However, the results deviate at lower temperatures for two reasons: first, a cooling-flow with large density perturbations, as is the case in our simulations, can give rise to blobs at $\sim 10^4$~K cooling out of the soft X-ray emitting gas, whereas a steady cooling-flow is much smoother~(see Appendix \ref{app} for more details); second, our resolution is not high enough to obtain robust statistics for plasma at $\sim 10^5-10^6$~K. A plasma blob cooling isobarically from $10^7$~K shrinks by a factor $\sim 10^8$ in volume by the time it reaches the stable phase.\footnote{For a steadily cooling blob $M/t_{\rm cool} \approx$ constant, irrespective of the temperature bin, where $M$ is the mass in a given temperature bin. Thus, the volume occupied by a cooling blob in an isobaric cooling-flow is reduced by $\sim T^3/\Lambda$.} Thus cold filaments are  not spatially well-resolved even in our highest resolution simulations! However, we are able to get reasonable statistics by combining simulation data over a long time interval. Very high resolution simulations of the core to quantitatively calculate emission in different temperature bins will be carried out in the future.

\subsection{Cool-Core Clusters in Thermal Balance}
\label{sec:fid}

\begin{figure*}
\includegraphics[scale=0.31]{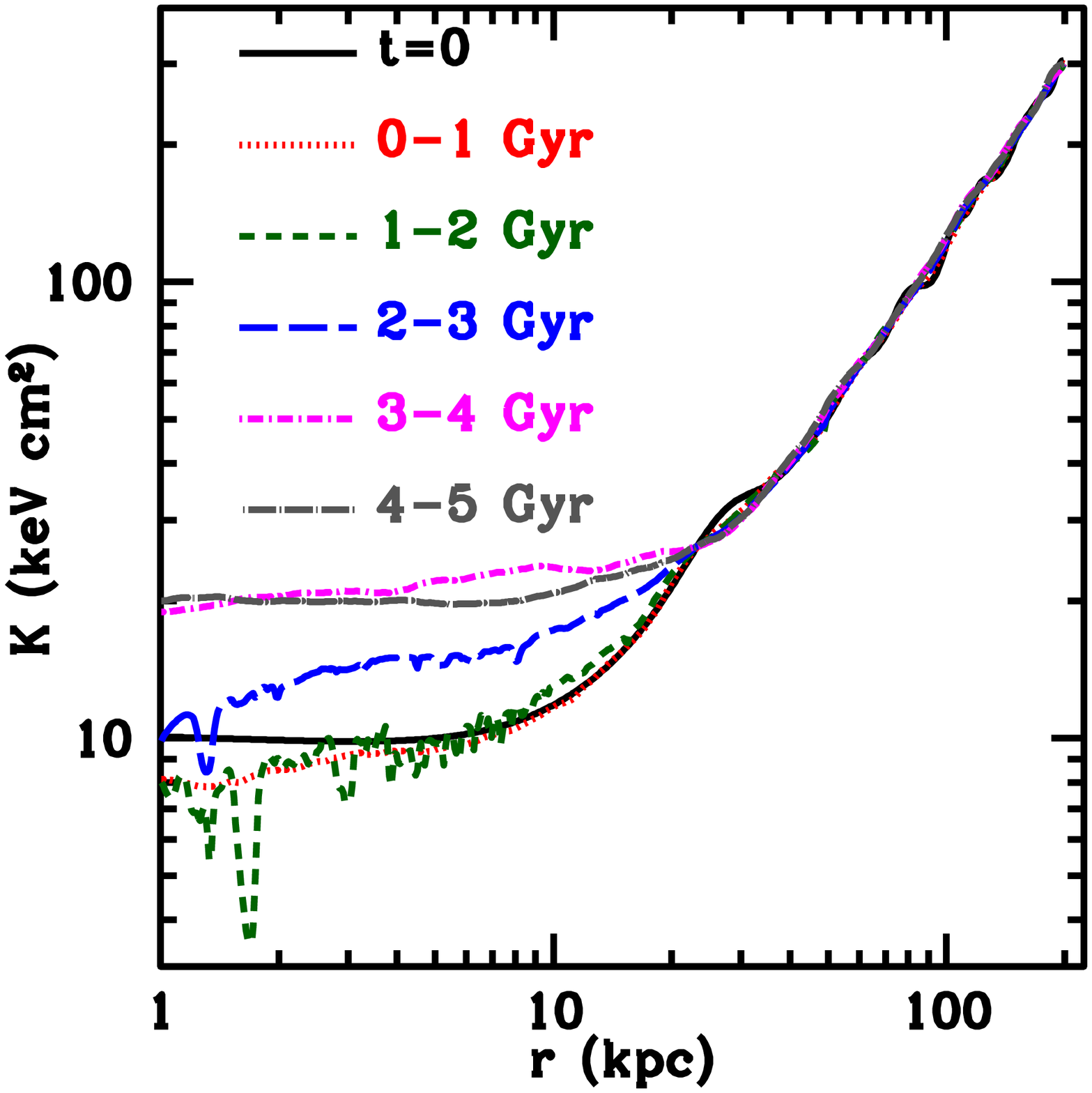} \includegraphics[scale=0.31]{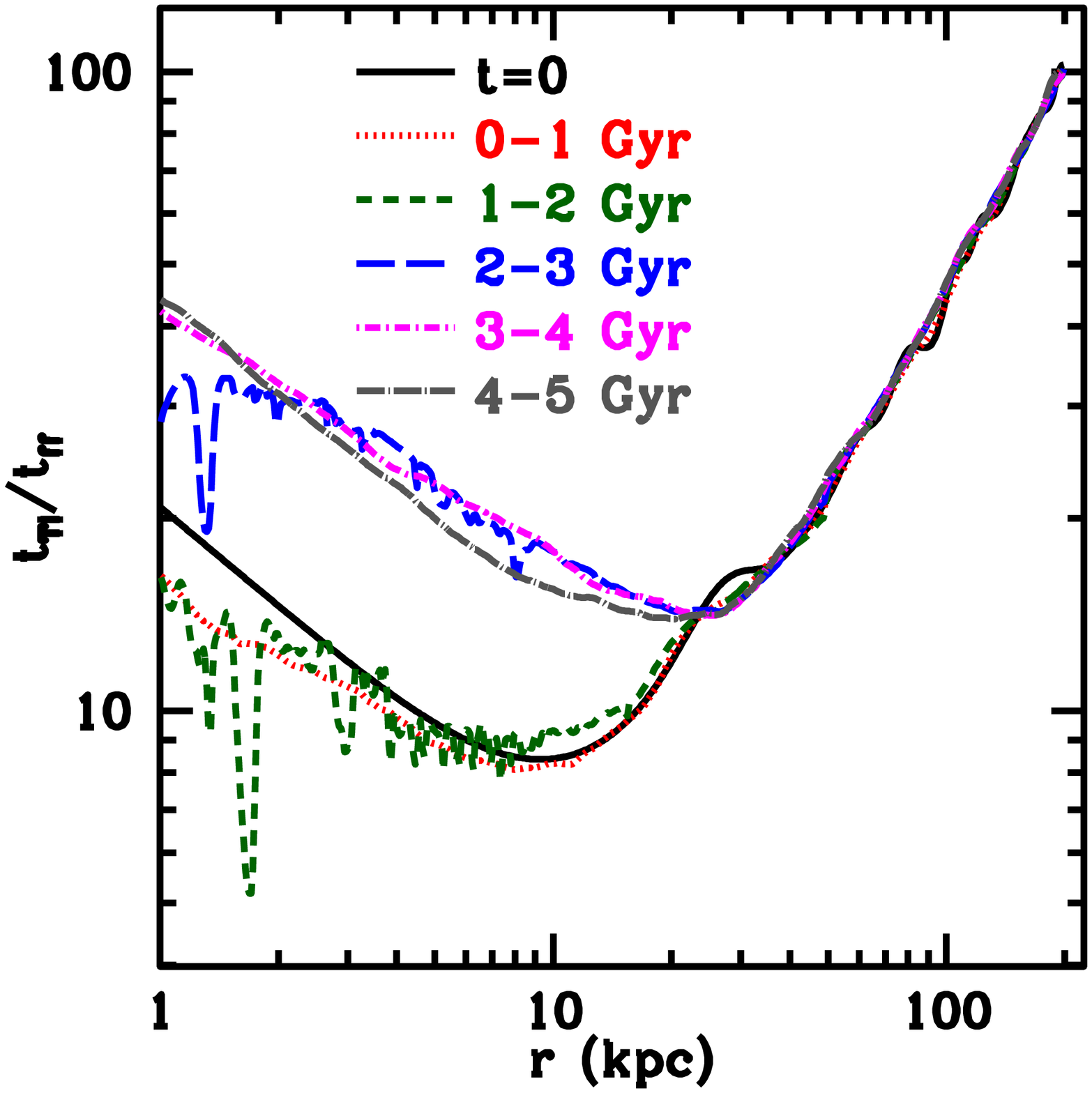}
\caption{Time- and angle-averaged entropy~({\em left}; $\langle n_eT \rangle/\langle n_e \rangle^{5/3} $, where $\langle \rangle$ represents averaging in $\theta$ and $\phi$) in $\kent$ and $t_{\rm TI}/t_{\rm ff}$~({\em right}) as a function of radius for the fiducial run C10. Solid black line represents the initial profile. The other lines are the profiles averaged over a 1~Gyr timescale at different times. Sharp low entropy~(and low $t_{\rm TI}/t_{\rm ff}$) troughs seen from 1--2~Gyr and 2--3~Gyr are because of infalling cold gas filaments, which are also responsible for  the large accretion rates seen in Figures~\ref{fig:mdot}~\&~\ref{fig:mdot_C10} during the same time intervals.
\label{fig:ent_C10}}
\end{figure*}

Figure~\ref{fig:2D} shows snapshots of the inner radii in our fiducial idealized heating run~(C10) at times when cold filaments condense out of the hot phase. The top-left panel shows a snapshot right before the condensation of a cold `filament.'  Once the over-dense blob cools below $\sim 1$~keV, the
cooling rate becomes faster than the turbulent mixing rate and the blob
quickly cools to the thermally stable temperature.  This newly born cold
filament has a non-zero velocity because of the motion of the overdense blob
from which it formed; cold filaments can thus have instantaneous velocities
directed away from the center of the potential. Further cooling of the hot phase close to the cold filament seeds condensation of secondary filaments. Since filaments are born with non-zero velocity, they move through the ICM core before leaving through the inner boundary after $\sim {\rm few}$ free-fall times. The cold blobs at the thermally stable temperature~($\sim 10^4$~K) are in rough pressure equilibrium with the hot phase. The dramatic difference in density~(and temperature) of the cold and hot phases, and the lack of plasma at intermediate densities~(temperatures) clearly points to local TI as the origin of these cold filaments. The production of cold filaments ceases once the density of the hot phase is depleted such that $t_{\rm TI}/t_{\rm ff} \gtrsim 10$; the bottom-right panel of Figure~\ref{fig:2D} shows the core in a quasi-steady state without filaments. 

Multiphase structure also develops in cooling-flows with {\em large} density perturbations because of a faster cooling rate in overdense blobs~(see Appendix \ref{app} for a detailed analysis). Whereas the formation of multi-phase gas in our simulations with
idealized heating depends only on the ratio $t_{\rm TI}/t_{\rm ff}$, production of multi-phase gas in cooling-flows depends also on the amplitude of the density perturbations.

Figure~\ref{fig:ent_C10} shows the time-(over 1~Gyr intervals) and angle-averaged profiles of entropy~(left panel) and the ratio $t_{\rm TI}/t_{\rm ff}$~(right panel) for the fiducial run from 0 to 5~Gyr. The cold filaments of the kind seen in Figure~\ref{fig:2D} show up as troughs in the entropy and $t_{\rm TI}/t_{\rm ff}$ profiles at 1--2 and 2--3~Gyr. Having condensed from the hot phase, these dense filaments fall through the inner boundary at roughly the free-fall rate. By 3~Gyr the hot ICM has both been heated and lost sufficient mass due to ``drop-out" of cold filaments that $t_{\rm TI}/t_{\rm ff} > 10$ everywhere. There are no more cold filaments produced once $t_{\rm TI}/t_{\rm ff} > 10$, even though the cooling time in the core is less than 1~Gyr. This is also reflected in the entropy profile as a rise in the central entropy to $K_0 \approx 20 \kent$ in the final state. As mentioned already, cold filaments condense out of the hot phase only when the ratio $t_{\rm TI}/t_{\rm ff} \lesssim 10$, or equivalently when the core entropy $K_0 \lesssim 20 \kent$.

\begin{figure}
\centering
\includegraphics[scale=0.42]{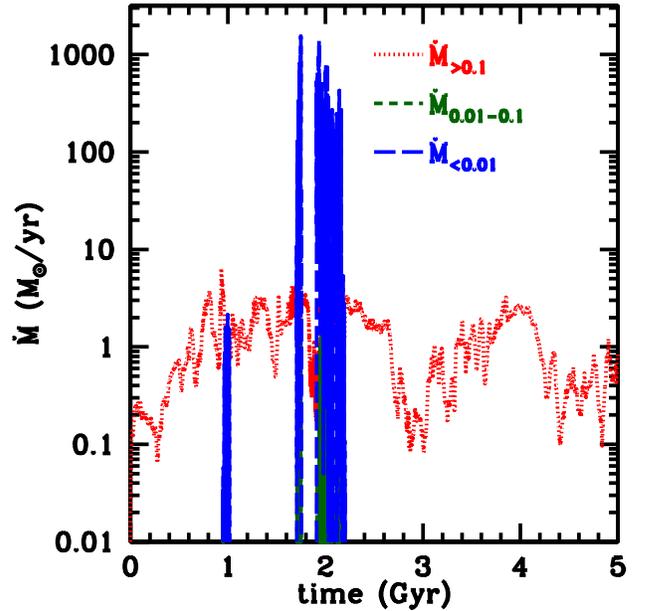}
\caption{Mass accretion rate through the inner boundary~(1~kpc) for the fiducial run~(C10) as a function of time. The  accretion rate in different temperature phases is indicated by different colors/styles. The instantaneous accretion rate in the cold~($T<0.01$~keV) phase can be large, reaching $\gtrsim$ 100 $\mpy$, but lasts only for short bursts~(see Fig.~\ref{fig:time_vs_mdot}). Most of the time accretion occurs in the hot~($T>$0.1~keV) phase. Accretion of plasma at intermediate temperatures~(0.01~keV $< T <$ 0.1~keV) coincides with the infalling cold filaments, but is subdominant. The green short-dashed line indicating accretion at intermediate temperatures is obscured by the blue long-dashed line showing accretion in the cold phase. 
\label{fig:mdot_C10}}
\end{figure}

The cold filaments falling toward the center cause a  sudden increase in the mass accretion rate at small radii. Figure~\ref{fig:mdot_C10} shows the mass accretion rate through the inner radius~(1~kpc) in different phases for the fiducial run: cold phase with $T<0.01$~keV, intermediate phase with 0.01~keV $<T<0.1$~keV, and hot phase with $T>0.1$~keV. These temperature ranges are chosen somewhat arbitrarily, but the results do not depend sensitively on the exact temperature values because there is very little plasma between the stable phase~(at $\sim 10^4$~K) and about one-third of the virial temperature (see section \ref{sec:comp}). The big spikes in $\dot{M}$ for the fiducial run C10 in Figure~\ref{fig:mdot} coincide with cold filaments~(seen in Fig.~\ref{fig:2D}) falling through the inner boundary. The mass accretion rate in the hot and intermediate phases is $< 10~\mpy$ at all times. However, the mass accretion rate in the cold phase can reach up to 1000 $\mpy$ for short times due to cold dense filaments crossing the inner boundary. The infalling cold filaments have a sheath of plasma at intermediate temperatures which contributes to the small mass flux at intermediate temperatures during the infall of cold filaments.

\subsection{Self-Regulation of the ICM core}
\label{sec:self_reg}

\begin{figure*}
\centering
\includegraphics[scale=0.31]{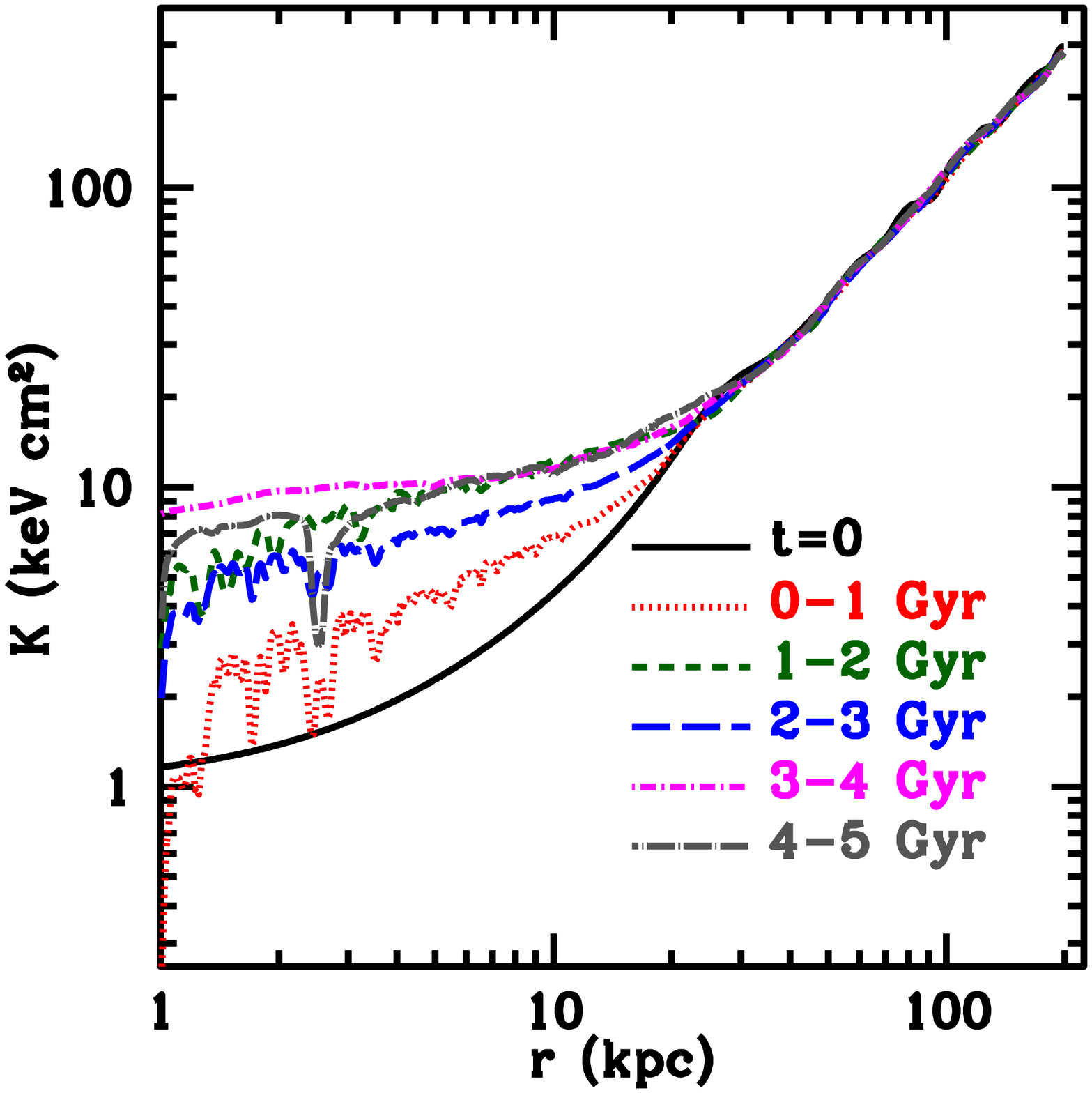} \includegraphics[scale=0.31]{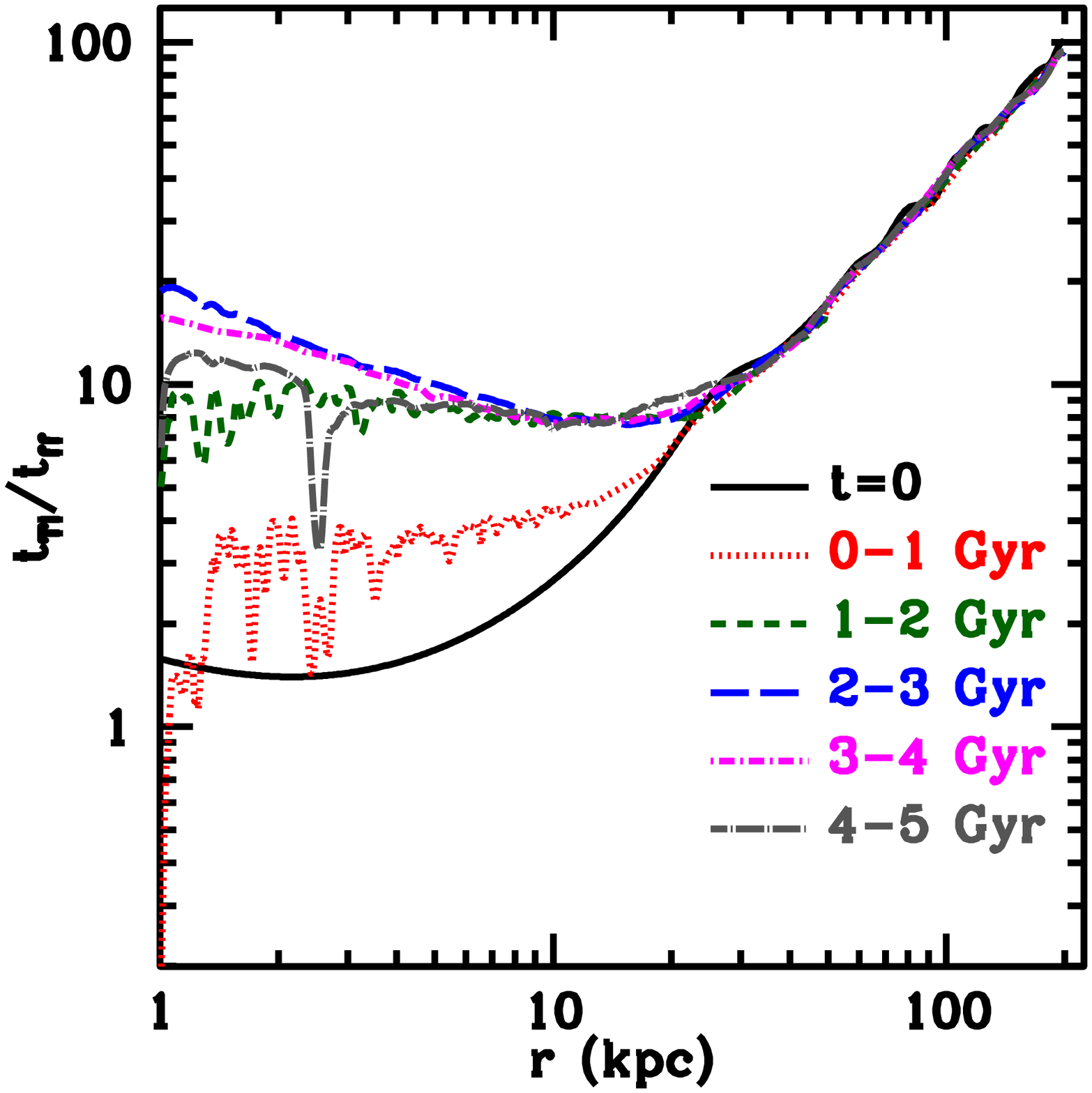}
\caption{Time- and angle-averaged entropy~({\em left}; $\langle n_eT \rangle/\langle n_e \rangle^{5/3} $) in $\kent$ and $t_{\rm TI}/t_{\rm ff}$~({\em right}) as a function of radius for the run C1. Solid black line represents the initial profile. The other 
lines are the profiles averaged over a 1~Gyr timescale at different times. 
\label{fig:ent_C1}}
\end{figure*}

Having discussed how the timescale ratio $t_{\rm TI}/t_{\rm ff}$ influences the accretion rate and the amount of cold gas in the ICM, we now discuss how these effects in turn influence the ratio $t_{\rm TI}/t_{\rm ff}$ in the cool-core of the ICM. We show in this section that the local TI tends to maintain clusters close to the critical threshold for forming filaments such that $t_{\rm TI}/t_{\rm ff} \sim 10$. The condition for forming cold filaments~($t_{\rm TI}/t_{\rm ff} \lesssim 10$) can be expressed in terms of entropy and the free-fall time as
\be
\label{eq:ent_th}
K \lesssim 20~\kent  \left [ T_{\rm keV}^{1/2} \Lambda_{-23} \left ( \frac{t_{\rm ff}}{30~{\rm Myr}} \right ) \right ]^{2/3},
\ee
where we have assumed a heating rate which is constant per unit volume~($q^+ \propto \rho^0$). Thus our thermal instability model can explain the observational lack of cold gas indicators in clusters with $K_0 > 30 \kent$ and the observed over-abundance of cool-core clusters at $K_0 \sim 10 \kent$~(e.g., Fig.~1 in \citealt{cav08}). 

\begin{figure}
\centering
\includegraphics[scale=0.42]{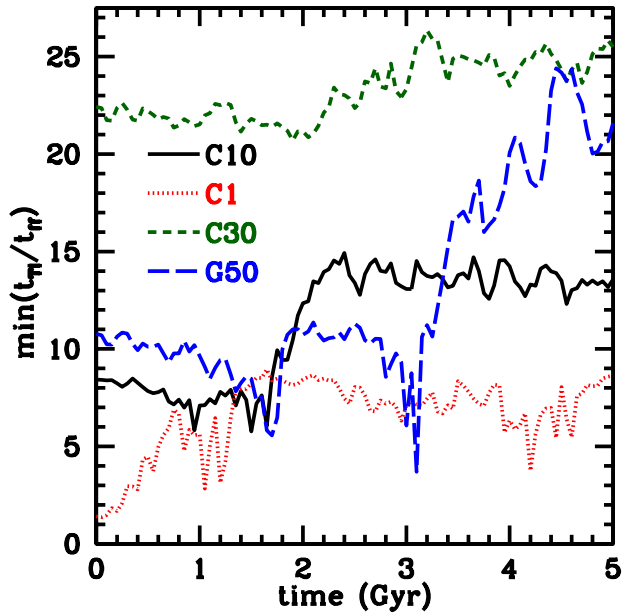}
\caption{The minimum of the ratio $t_{\rm TI}/t_{\rm ff}$ as a function of time for various runs~(evaluated from angle-averaged radial profiles of plasma with temperature $>$ 0.1~keV sampled every 50~Myr). Cold filaments condense out and result in large accretion rates whenever the ratio $t_{\rm TI}/t_{\rm ff} \lesssim 10$~(compare with the left panel of Fig.~\ref{fig:mdot}). This in turn leads to significant heating, driving $t_{\rm TI}/t_{\rm ff}$ to be $\gtrsim 10$. 
\label{fig:trat_min_1}}
\end{figure}

Figure~\ref{fig:ent_C1} shows the entropy and $t_{\rm TI}/t_{\rm ff}$ profiles at different times for a cluster with an initially low entropy core C1~($K_0=1 \kent$). Cold filaments quickly condense out of the hot phase since the ratio 
$t_{\rm TI}/t_{\rm ff} < 10$ everywhere within 20~kpc. The infalling cold filaments leave the inner boundary after roughly a free-fall time. 
The entropy of the remaining plasma, maintained in hydrostatic equilibrium by our idealized heating prescription, thus increases. 
For extremely low-entropy initial conditions the central entropy increases very quickly because of a short cooling time; this happens in less than 1~Gyr for the run C1~(see Fig.~\ref{fig:trat_min_1}). After a few TI times, the density profile~(and the entropy profile as well) adjusts such that the ratio $t_{\rm TI}/t_{\rm ff}$ is close to the critical value of $\approx 10$. This adjustment of the inner ICM because of TI and infalling cold filaments plays a very important role in determining the density profile of the hot plasma in halos of different masses, as we discuss further in section \ref{sec:diff_mass}.

The average entropy at small radii from 4--5~Gyr for the run C1 is slightly lower than the inner entropy at late times for the fiducial run~(see Fig.~\ref{fig:ent_C10}). This slightly lower entropy~(and $t_{\rm TI}/t_{\rm ff}$ smaller than the critical value) results in another episode of filaments forming and falling through the inner boundary at 4~Gyr, as seen in the left panel of Figure~\ref{fig:mdot}. Thus, $t_{\rm TI}/t_{\rm ff} \lesssim 10$ is a robust criterion for the presence of filaments, irrespective of the initial conditions. Moreover, the critical ratio $t_{\rm TI}/t_{\rm ff}$ is self-regulated to be $\gtrsim 5-10$ because of heating and mass drop-out in the form of cold filaments. Also notice that the inner core at late times does not have a flat entropy profile, though the profile is shallower compared to the larger radii. While the exact entropy profile at small radii depends on the details of heating and cooling, an isentropic core appears to be a good approximation because the cool low-entropy plasma blobs leave from the inner boundary and the hot plasma is mixed throughout the core.

Figure~\ref{fig:trat_min_1} shows the minimum of the ratio $t_{\rm TI}/t_{\rm ff}$ as a function of time for the cluster runs C1, C10, and C30~(and the group run G50 discussed in section \ref{sec:diff_mass}). As can be seen in Figures~\ref{fig:ent_C10}~\&~\ref{fig:ent_C1}, the minimum in $t_{\rm TI}/t_{\rm ff}$ increases due to mass drop-out in the form of cold filaments for runs C1 and C10. Whenever $t_{\rm TI}/t_{\rm ff} \lesssim 10$ in Figure~\ref{fig:trat_min_1}, cold filaments condense out of the hot phase and fall in through the inner boundary, and are seen as spikes in the mass accretion rate in the left panel of Figure~\ref{fig:mdot}. The minimum of the ratio $t_{\rm TI}/t_{\rm ff}$ is larger than the critical value~(10) for the run C30~(with $K_0=30\kent$) and there are no cold filaments~(e.g., see Table \ref{tab:tab1}). Figure~\ref{fig:trat_min_1} shows that this ratio~(and the core entropy $K_0$) does not evolve much with time, even though the cooling time is $\approx$ 1~Gyr. The mass accretion rate for this run is much smaller than for the cooler clusters because massive cold filaments cannot condense out of the hot phase~(see left panel of Fig.~\ref{fig:mdot}).  Although the cooling time is less than the cluster age for run C30 there is no cooling-flow because of our heating prescription~(compare C30 and C30-cf in Fig.~\ref{fig:mdot}). Thus, even clusters which are unlikely to show cold filaments~(since $t_{\rm TI}/t_{\rm ff} > 10$), but which have cooling times shorter than the cluster age, require heating roughly balancing cooling to prevent a massive cooling-flow. 

The rapid adjustment of cool-core clusters to $t_{\rm TI}/t_{\rm ff} \gtrsim 10$ via mass drop-out and heating, and the suppression of $\dot{M}$~(and hence feedback) if $t_{\rm TI}/t_{\rm ff} \gtrsim 10$ suggest that cool-core clusters will spend most of their time close to the threshold for forming cold filaments, so long as heating is related to the amount of cold gas produced by TI~(as is the case for AGN/supernova feedback). This physics can quantitatively account for the observed excess of clusters with central entropies around $10 \kent$~(e.g., \citealt{cav08,pra10}).  In addition to cool-core clusters affected by cooling and feedback, there are also non-cool-core clusters which have high entropies because of strong feedback heating and/or mergers. These non-cool-core clusters, relatively unaffected by cooling, maintain their high entropies and form a class distinct from the cool-core clusters.

\subsection{Comparison of Different Halo Masses}

\label{sec:diff_mass}

We have also carried out simulations with our idealized heating prescription
in lower mass, group-sized halos~($M_0=3.8 \times 10^{13} \msun$), in massive groups with $M_0=10^{14} \msun$, and in more
massive cluster halos~($M_0=10^{15}$)---these appear in Table~\ref{tab:tab1} with
the prefixes `G,' `MG,' and `MC,' respectively.  The number after the prefix represents
the entropy of the equivalent fiducial-mass halo, assuming self-similar
scaling.  Thus, runs G300 and C300 have identical density profiles, but the
temperature in run G300 is lower by a factor of $10^{2/3}$ and the core
entropy in G300 is $K_0=300/10^{2/3}=64.6\kent$.

At a given scaled entropy, cooling becomes stronger in lower mass halos
because of the corresponding decrease in the virial temperature~($t_{\rm cool}
\propto T^{1/2}/n \propto M^{1/3}$).  On the other hand, the free-fall time
($t_{\rm ff} \propto [r^3/M_{\rm enc}]^{1/2}$) at a given~(scaled) radius is
roughly independent of the halo mass because dark matter, which dominates
gravity, is self-similar.  Consequently, groups develop multi-phase gas at a
higher scaled entropy than do clusters; e.g., while cold filaments do not
condense from the hot phase in the cluster run C30, filaments are seen in the
group run G50.

The group run G50 is, in fact, initially very close to the threshold for
forming multi-phase gas.  Figure~\ref{fig:trat_min_1} shows the minimum of the
ratio $t_{\rm TI}/t_{\rm ff}$ as a function of time for this simulation.  It
does not show cold filaments from 0--1~Gyr, but because of slightly low entropy
material falling in from larger radii, it crosses the threshold for forming
cold filaments after 1~Gyr.  The group also undergoes a second, stronger
episode of thermal instability and again forms filaments after $\approx 3$ Gyr;
this depletes the core of much of its gas and raises the entropy in the hot
phase much beyond the critical threshold.

Figure~\ref{fig:time_vs_mdot} shows that the mass accretion rate in the hot
phase is smaller in groups than in clusters by a factor of $\sim 10$~(their
mass ratio; see also Table \ref{tab:tab1}), but that the mass accretion rate
in the cold phase~(which accounts for any $\dot{M} \gtrsim {\rm few}\mpy$) is
nearly independent of the halo mass.  Thus, the amount of gas which reaches the
cold phase is similar in our group- and cluster-mass halos and these halos
therefore receive similar amounts of feedback energy.  The mass of the hot
phase is lower in groups than in clusters, however, and the effect of this
feedback is therefore much stronger.  Thus, even in our idealized heating
model, heating of the hot phase can overwhelm cooling in lower mass halos,
resulting in a much lower density~(and larger $t_{\rm TI}/t_{\rm ff}$) than
the critical threshold for filament formation.  The last 2~Gyr in
Figure~\ref{fig:trat_min_1} illustrate this overshoot.

\subsubsection{Bigger Cool-Cores for Lower Mass Halos}

\begin{figure}
\centering
\includegraphics[scale=0.42]{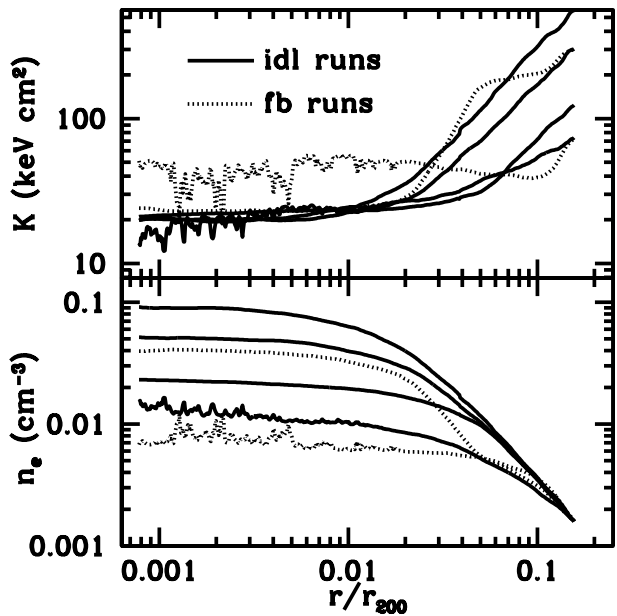}
\caption{The angle- and time-averaged~(4--5~Gyr) electron number density~($n_e~{\rm in cm}^{-3}$) and entropy~(keV~cm$^2$) as a function of radius~(scaled by the virial radius) for the idealized heating runs~(MC1, C10, MG10, \& G50;  solid lines) and for the runs with feedback heating~(C10-fb-2em3 with $\epsilon_{\rm fb}=2\times10^{-3}$ \& G50-fb-7em5 with $\epsilon_{\rm fb}=7\times10^{-5}$; dotted lines). Only the hot X-ray emitting plasma with $T>$ 0.1~keV is used to calculate the average radial profiles. Both the core density and temperature decrease with a decreasing halo mass; therefore the core entropy is relatively independent of the halo mass. 
\label{fig:nTvsr_sims}}
\end{figure}

Figure~\ref{fig:nTvsr_sims} shows the angle averaged entropy and electron number density in the hot phase~($T>0.1$~keV; time averaged from 4--5~Gyr) as a function of radius~(scaled to the virial radius) for simulations with different halo masses; only halos prone to the formation of cold filaments according to our $t_{\rm TI}/t_{\rm ff}$ criterion ~(runs MC1, C10, MG10, \& G50) are shown. The ICM with $t_{\rm TI}/t_{\rm ff} \lesssim 10$ develops cold filaments which remove mass, and bring $t_{\rm TI}/t_{\rm ff}$, entropy, and density in the core close to the critical value.  Thus, the core density~(entropy) for each halo in Figure~\ref{fig:nTvsr_sims} represents a limit below~(above) which cold filaments cannot condense out for a given halo mass. While the temperature decreases with a decreasing halo mass, the core radius~(where the ratio $t_{\rm TI}/t_{\rm ff} \propto T^{1/2}/(n t_{\rm ff})$ is the smallest) increases, assuming a similar density profile outside the core radius. Thus the free-fall time $t_{\rm ff} \propto~(r/r_{200})^{1/2}$ at the core radius increases with a decreasing halo mass. These trends in temperature and $t_{\rm ff}$ at the core radius roughly cancel, leading to an entropy threshold for filament formation~(Eq. \ref{eq:ent_th}) that is similar for groups and clusters. This can be verified from Figure~\ref{fig:nTvsr_sims} which shows a relatively constant entropy for different halo masses; as a result, the core entropy for groups is larger than the self-similar prediction $K\propto M_0^{2/3}$, as is observed~(e.g., \citealt{pon99,sun09b}).

\subsubsection{Lack of Gas below $T_0/3$}

\begin{figure}
\centering
\includegraphics[scale=0.42]{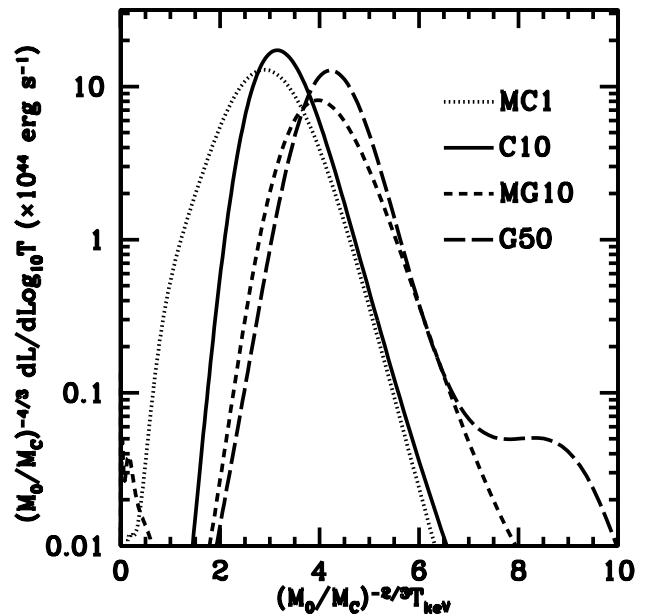}
\caption{Luminosity per logarithmic temperature bin~($dL/{\rm Log}_{10}T$) as a function of temperature for different halo masses; the data are averaged over a 1~Gyr interval when the models have cold filaments. The axes are scaled according to the self-similar prediction for luminosity~($\propto M_0^{4/3}$) and temperature~($\propto M_0^{2/3}$). The normalization of both axes is relative to the cluster mass halo~($M_C=3.8 \times 10^{14} \msun$). 
\label{fig:lum_vs_T_Mh}}
\end{figure}

In section \ref{sec:subv} we showed that for our cluster model in thermal balance only a tiny fraction of the plasma cools below $\approx 1$~keV. Here we discuss the cooling of the hot gas in different halo masses. Figure~\ref{fig:lum_vs_T_Mh} shows the luminosity~($dL/d\log_{\rm 10}T$) as a function of temperature for simulations with different halo masses which show cool filaments~(MC1, C10, MG10, \& G50; these are all `cool-core' halos). The axes are scaled according to the self-similar prediction. The profiles are obtained by averaging over 1~Gyr coinciding with the last episode of condensation due to TI. For all halos the luminosity drops quite abruptly below the temperature corresponding to the peak luminosity. Moreover, the shape of $dL/d{\rm Log}_{10}T$ vs. temperature below the peak temperature is quite similar, and is consistent with the observational fact that there is negligible luminosity below $1/2-1/3$ of the ambient cluster temperature \citep[e.g.,][]{pet03,san10}. The line corresponding to the group G50 in Figure~\ref{fig:lum_vs_T_Mh} shows a  small bump at double the ambient temperature because some of the gas is heated above the virial temperature; this is consistent with the late-time rise in the ratio $t_{\rm TI}/t_{\rm ff}$ for G50 in Figure~\ref{fig:trat_min_1}.

\subsubsection{The Required Heating Efficiency}
\label{sec:req_eff}
The last columns of Tables \ref{tab:tab1} \& \ref{tab:tab2} list the required average heating efficiency $\epsilon = Q^+/\dot{M}c^2$ corresponding to our heating 
prescription, where $Q^+ = \int_{r_{\rm in}}^{r_{\rm cool}} q^+(r)4\pi r^2dr$~($q^+$ is given by Eq. \ref{eq:heat}), $r_{\rm in}$ is the inner radius, $r_{\rm cool}$ is the cooling radius where the angle-averaged cooling time is $\approx 5$~Gyr, and $\dot{M}$ is the average accretion rate through the inner radius. In calculating the efficiency we do not integrate over the whole domain, although we do apply our heating prescription at all radii, because the radii where the cooling time is longer than 5~Gyr do not require heating to maintain the pressure support. Thus, the efficiency is zero for the high entropy runs with cooling times longer than 5~Gyr. 
Among cluster runs~(prefixed by C), the required efficiency~($\epsilon$) peaks at $\approx 5 \times 10^{-3}$ for the lowest core entropy run which does not show cold filaments~(run C30). Even for other halo masses, the required heating efficiency roughly peaks for the lowest entropy run not showing filaments. The required efficiency is smaller once cold filaments can condense because while the global cooling rate is similar to the higher entropy runs with no filaments, the mass accretion rate is dramatically larger.

The required heating efficiency for runs at the threshold for forming cold filaments~(MC1, C10, MG10, and G50) decreases with a decreasing halo mass; e.g., while $\epsilon$ for the run C10 is  $10^{-3}$, it is  $7.4 \times 10^{-5}$ for G50, more than ten times smaller. This is because the mass in the hot phase decreases by $\approx 10$ from C10 to G50, but the mass accretion rate~(primarily in the cold phase) is about the same~(Fig.~\ref{fig:time_vs_mdot}; see also Table \ref{tab:tab1}). Thus, a smaller feedback efficiency is sufficient to balance cooling for a smaller mass halo. For a realistic feedback scenario this means that while AGN feedback is required for group and cluster mass halos~(supernova feedback is quite inefficient, $\epsilon \lesssim 10^{-5}$), supernova feedback may be sufficient for Galactic scale halos that have an even smaller mass in the hot phase. A similar conclusion is reached for our simulations with feedback heating discussed in section \ref{sec:fb}. One implication of this result is that, if AGN heating with a fixed feedback efficiency is responsible for heating cool cores, the lower mass halos can be heated much more than more massive ones.

\subsection{Robustness of the Idealized Heating Ansatz}
\label{sec:robust}

\subsubsection{Spatial \& Temporal Perturbations on Heating}

To test the robustness of our results we also carried out simulations with spatial and temporal perturbations in heating on top of thermal balance at every radius. Simulations with rather large variations about this balance are qualitatively similar to the idealized simulations~(Table \ref{tab:tab2}; see also \citetalias{mcc11}). The run C10-p10 is similar to the fiducial case but with temporal perturbations  in the heating rate of a factor of $\approx 10$ at each grid cell~(on top of $q^+$ given by Eq. \ref{eq:heat}). The perturbations, chosen from a uniform distribution centered at zero, have a correlation time of 100~Myr~(the results do not depend sensitively on the correlation time). The entropy and $t_{\rm TI}/t_{\rm ff}$ profiles are similar to those for the fiducial run in Figure~\ref{fig:ent_C10}. Table \ref{tab:tab2} also shows that the higher entropy run with perturbations C30-p10 does not show cold filaments, consistent with our idealized run C30. Thus, our picture of TI holds even with large perturbations in the heating rate because on average the plasma is still in thermal equilibrium.

We also carried out simulations~(C10-nz1, C10-nz4, etc.) in which the balance of heating and cooling is imposed on larger scales, rather than at every radius as is the case in the idealized simulations we have focused on; e.g., in run C10-nz16 the heating rate in each of 16 logarithmically spaced radial zones is calculated by averaging cooling over each zone. From these simulations we find that if heating balances cooling on scales smaller than the cooling scale height~(the scale over which the cooling rate changes significantly) then the results are similar to the idealized heating runs.
We thus conclude that, as argued in \citetalias{mcc11}, many of 
the details of heating are not important for the thermal evolution of the halo, as long as the heating stabilizes the ICM against a cooling catastrophe.

\subsubsection{Three-Dimensional Runs \& Convergence Studies}

Table \ref{tab:tab2} lists three-dimensional~($256 \times 128 \times 32$) simulations of clusters C10 and C30. The two- and three-dimensional simulations agree in that the runs C30 and C30-3D with $t_{\rm TI}/t_{\rm ff} > 10$ do not show cold filaments but the runs with the ratio $t_{\rm TI}/t_{\rm ff} < 10$~(C10 and C10-3D) show filaments. The mass in the cold phase, mass accretion rate, etc. are not identical  in two and three dimensions because there are non-axisymmetric initial perturbations in the three-dimensional simulations.

Table \ref{tab:tab2} also lists two-dimensional simulations but with higher and lower resolution than the fiducial case. The lower core entropy~(C10, C10-hlf, C10-dbl) cases show cold filaments for all resolutions. Similarly, the higher core entropy runs~(C30, C30-hlf, C30-dbl) do not show cold filaments for any resolution. The threshold for forming multi-phase gas and the integrated quantities such as the average mass accretion rate are roughly independent of the resolution. We showed in \citet{sha10} that thermal conduction must be included in order to achieve strict convergence. Since we cannot resolve the Field length in the cold phase with a realistic cooling function using present computational resources, we necessarily sacrifice strict convergence. 

\section{Simulations with Feedback Heating}
\label{sec:fb}

In this section we describe the feedback simulations in which the integrated~(isotropic) heating rate is proportional to the instantaneous mass accretion rate at the innermost radius. This feedback model is clearly only a rough approximation to heating by a central AGN, but as we shall show, it provides a useful demonstration that many of the results in the previous section are robust and do not depend sensitively on the precise details of the heating. Although the exact mechanisms which heat the ICM are not fully  understood, possible scenarios involve jets and bubbles launched by the central AGN. These processes are ultimately fed by accretion, however, and their power is likely to be proportional to the mass accretion rate. Feedback heating by jets, bubbles, or star formation likely lags the instantaneous accretion rate by up to $\sim 100$~Myr, the dynamical time at the core radius. Moreover, the feedback efficiency is likely to be time variable and may depend on the mass accretion rate. More realistic modeling of accretion and launching of jets is outside the scope of this paper and will be investigated in the future.

Our feedback simulations are initialized far from equilibrium, with the heating rate $q^+=0$ but the cooling rate $q^- > 0$. Nonetheless, after a transient phase, they are able to reach an approximate thermal equilibrium where average heating in the core roughly balances average cooling. Thus, the results from our idealized heating simulations  carry over to these more realistic feedback simulations. Table \ref{tab:tab3} lists the runs with feedback heating using $q^+$ in Eq. \ref{eq:heat_fb}. In section \ref{sec:fb_cl} we describe the feedback simulations for clusters. In section \ref{sec:fb_gp} we show that feedback simulations for groups require lower feedback efficiency to achieve thermal balance. Section \ref{sec:comp} compares the results from our idealized heating simulations discussed in the previous section with the feedback runs.

\subsection{Feedback Heating in Clusters}
\label{sec:fb_cl}

\begin{figure}
\centering
\includegraphics[scale=0.42]{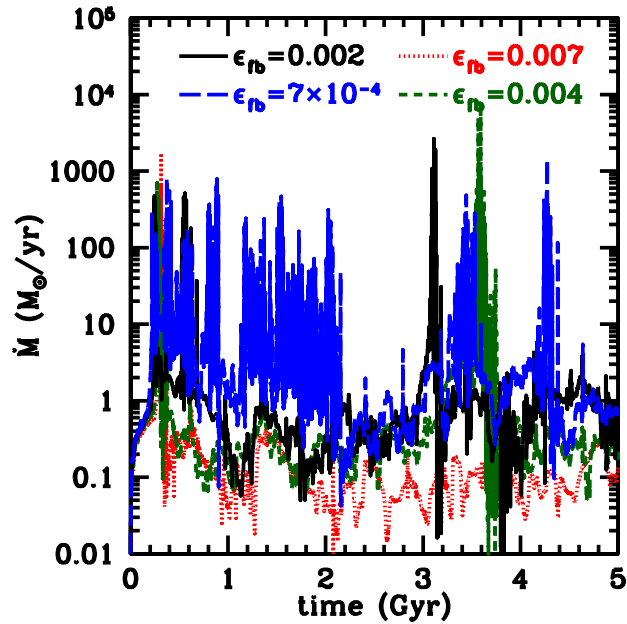}
\caption{Total mass accretion rate~(in $\mpy$) through the inner radius~(1~kpc) for the cluster runs with feedback heating. Different feedback efficiencies are labeled. Large spikes in the accretion rate correspond to infalling cold filaments crossing the inner boundary. The duty cycle for cold filaments and AGN feedback is shorter for a smaller feedback efficiency.
\label{fig:mdot_C10_fb}}
\end{figure}

Figure~\ref{fig:mdot_C10_fb} shows the mass accretion rate as a function of
time for the cluster simulation C10 with our feedback heating prescription.
We show simulations with feedback efficiencies $\epsilon_{\rm fb}$ = 0.007,
0.004, 0.002, $7\times 10^{-4}$~(see Eq.~\ref{eq:heat_fb}).  As discussed
above, all simulations are initially out of thermal equilibrium and the
accretion rates spike around 0.2~Gyr, or roughly the cooling time in the core.
If the feedback efficiency is large enough~(e.g., $\epsilon_{\rm fb}=0.007$),
the ICM approaches a steady-state with a low $\dot{M}$ and a low core density.
The ratio of the crucial time-scales $t_{\rm TI}/t_{\rm ff}$ is greater than $10$ at all
radii in this state, and cold filaments therefore do not form at later times.
If the efficiency is too low~(e.g., $\epsilon_{\rm fb} = 7 \times 10^{-4}$),
however, a larger mass flux is required to maintain thermal equilibrium in the
core; such high accretion rates can only be supplied by cold filaments.  The
evolution of these simulations is therefore sporadic, punctuated by episodes of
strong cooling or heating.  For intermediate feedback efficiencies
($\epsilon_{\rm fb} = 0.002,~0.004$), the ratio of time-scales $t_{\rm
TI}/t_{\rm ff}$ is typically larger than 10, but occasionally drifts to lower
values.  Filaments in these simulations are a rare, transient phenomenon.

In all cases, cold gas condenses from the ICM when~(and only when) the ratio
$t_{\rm TI}/t_{\rm ff} \lesssim 10$.  These infalling cold filaments drive
strong feedback and can overheat the ICM, as we discuss below.  The feedback
efficiency sets the duty cycle for the production of the multi-phase gas and
for feedback heating~(the duty cycle is shorter for a smaller feedback
efficiency), but not the critical threshold for the TI.  The excursions about
the critical threshold can be large, however, and depend on the feedback
efficiency.  The results are not sensitive to the initial core entropy and the
initial transients are wiped out on the core cooling timescale.

\begin{figure*}
\includegraphics[scale=0.31]{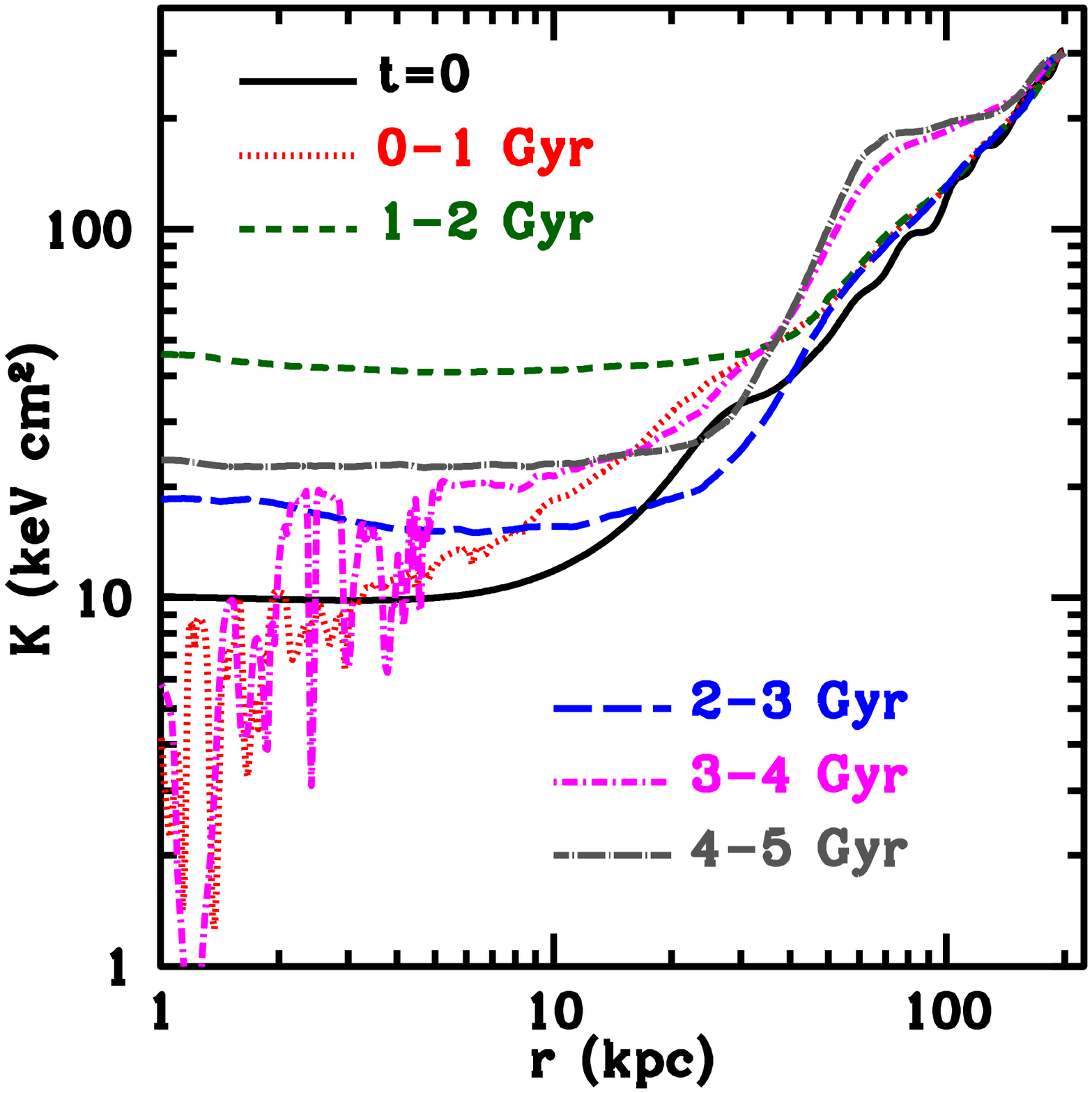} \includegraphics[scale=0.31]{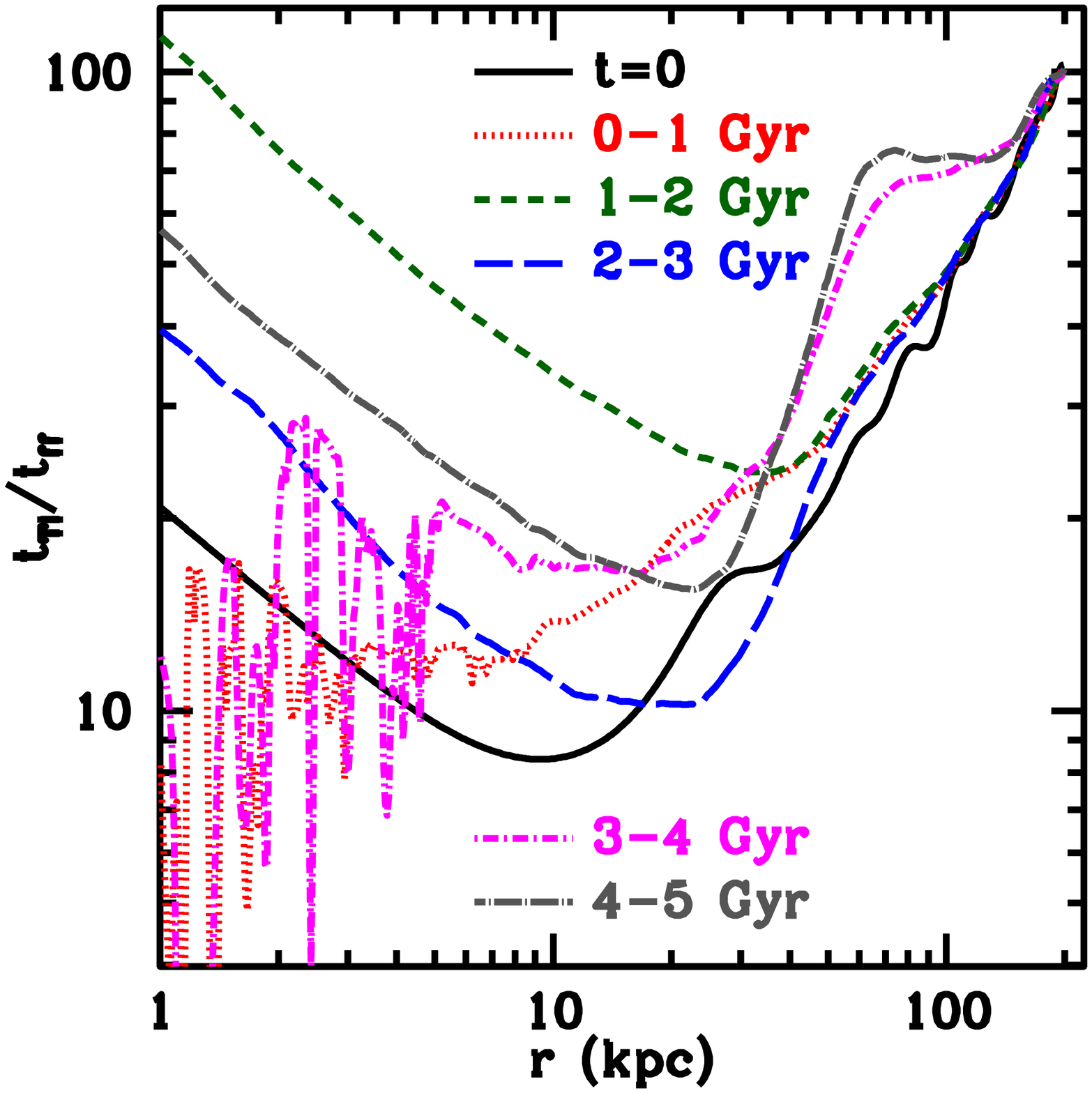}
\caption{Time- and angle-averaged entropy~({\em left}; $\langle n_eT \rangle/\langle n_e \rangle^{5/3} $) in $\kent$ and $t_{\rm TI}/t_{\rm ff}$~({\em right}) as a function of radius for the cluster run with feedback heating~($\epsilon_{\rm fb}=0.002$; run C10-fb-2em3 in Table \ref{tab:tab3}). Solid black line represents the initial profile. The other  lines represent the profiles averaged over a 1~Gyr timescale at different times. 
\label{fig:ent_C10_fb}}
\end{figure*}

Figure~\ref{fig:ent_C10_fb} shows profiles of entropy and of the ratio $t_{\rm
TI}/t_{\rm ff}$ for a simulation with an intermediate efficiency
$\epsilon_{\rm fb}=0.002$~(C10-fb-2em3).  Cooling initially dominates heating
and causes the ratio $t_{\rm TI}/t_{\rm ff}$ to drop below the critical
threshold of $\sim~10$; the TI thus creates cold filaments which reach the
center around 0.2~Gyr and lead to feedback heating.  Heating continues until
$t_{\rm TI}/t_{\rm ff}> 10$ everywhere and, by 1--2~Gyr, the core entropy has
significantly increased.  With this intermediate efficiency, however, the
hot-phase accretion is too small for the feedback to completely balance
cooling.  The core cools slowly but secularly and again crosses the $t_{\rm
TI}/t_{\rm ff} \lesssim 10$ threshold after another 1--2~Gyr.  This leads to a
second episode of accretion and heating.  This pattern---with the ICM maintaining 
rough thermal balance and producing multiphase gas whenever $t_{\rm TI} 
\lesssim 5-10~t_{\rm ff}$---would likely continue if we ran the simulation further.

The bump around 60~kpc in the profiles in Figure~\ref{fig:ent_C10_fb} result
from the particular choice of the heating parameters in Eq.~\ref{eq:heat_fb}.
Our aim here is not to produce a profile which perfectly matches observations
by varying all the parameters in the heating function, but rather to obtain
something reasonable without much fine tuning.

\subsection{Feedback Heating in Groups}
\label{sec:fb_gp}

\begin{figure}
\centering
\includegraphics[scale=0.42]{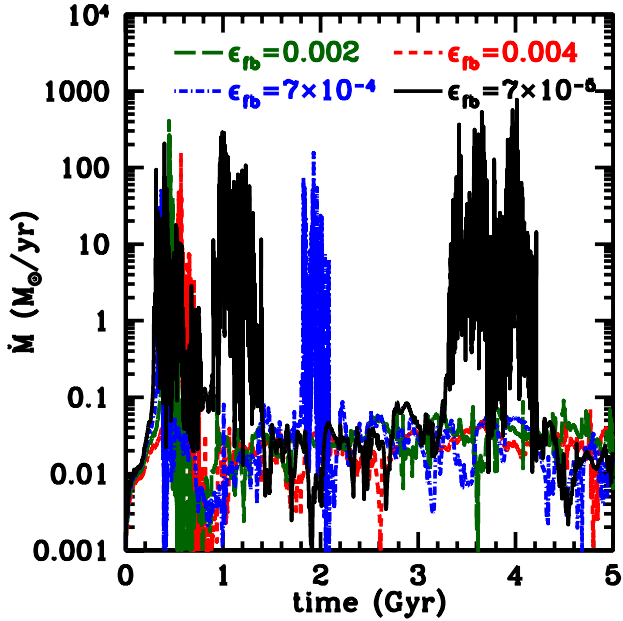}
\caption{Total mass accretion rate~(in $\mpy$) through the inner radius for group models~(G50) with feedback heating. Different feedback efficiencies are labeled. Large spikes in accretion correspond to infalling cold filaments crossing the inner boundary.
\label{fig:mdot_G50_fb}}
\end{figure}

Figure~\ref{fig:mdot_G50_fb} shows the mass accretion rate through the inner radius as a function of time for group mass halos~($M_0=3.8\times 10^{13} \msun$) with different feedback efficiencies. The cutoff radius~($r_{\rm cf}$) in Eq. \ref{eq:heat_fb} is chosen to be 70~kpc. Other parameters in the heating function~(Eq. \ref{eq:heat_fb}) are chosen to be the same as before~($\alpha=-1.5$ and $\beta=3.5$). The qualitative behavior of $\dot{M}$ as a function of time is similar to the clusters in Figure~\ref{fig:mdot_C10_fb} in that accretion in the cold phase becomes progressively more important as the feedback efficiency is reduced. Notice, however,
that the transition from a large accretion rate in the cold phase to a state with accretion primarily in the hot phase takes place at a lower feedback efficiency. For clusters this transition happens around $\epsilon_{\rm fb} \approx 0.002$ but in groups it occurs for $\epsilon_{\rm fb} \approx 7 \times 10^{-5}$. This conclusion was also reached for simulations with idealized heating and has implications for possible sources of feedback in groups/clusters and galactic halos.

\begin{figure*}
\centering
\includegraphics[scale=0.31]{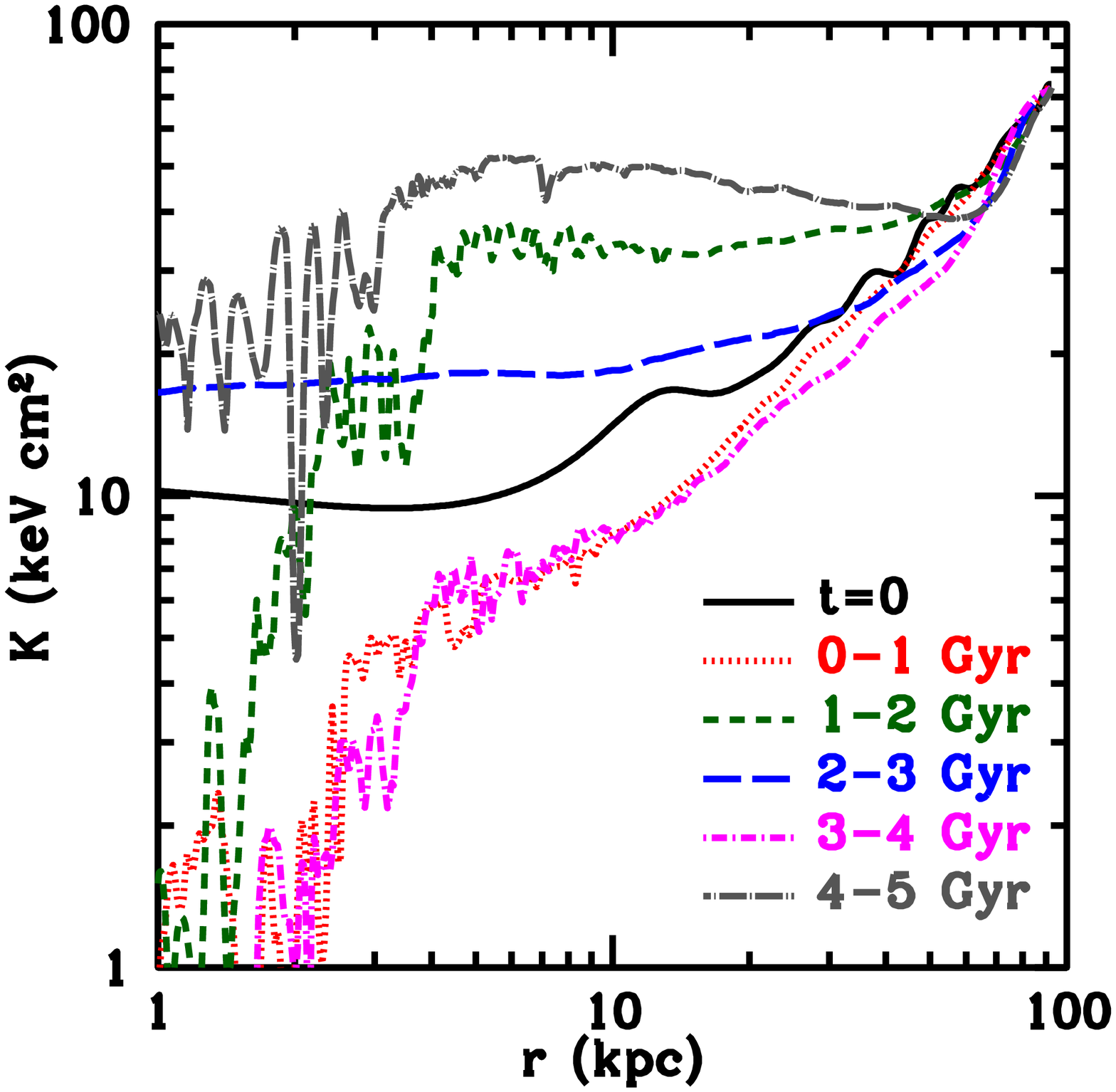} \includegraphics[scale=0.31]{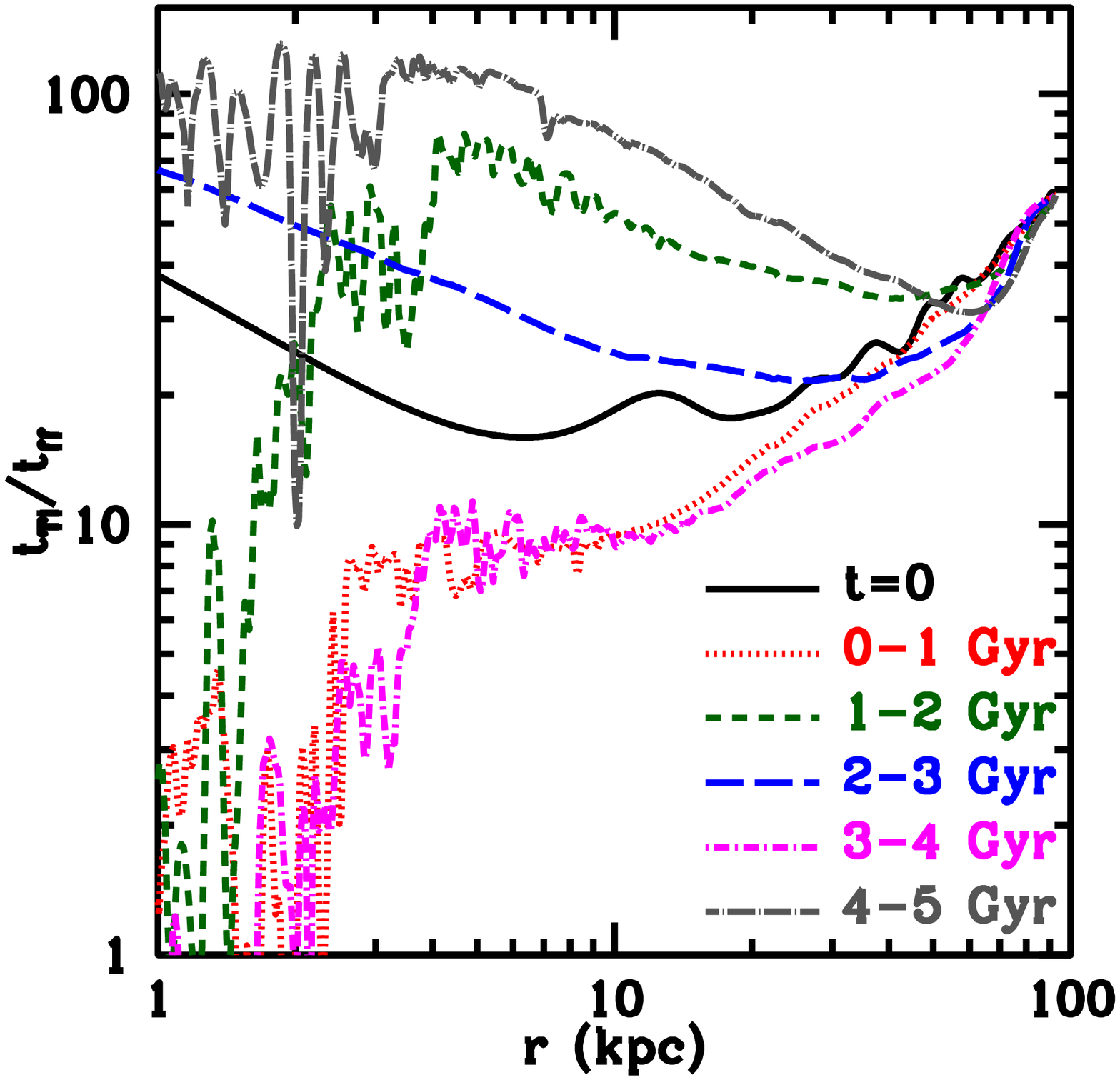}
\caption{Time- and angle-averaged entropy~({\em left}; $\langle n_eT \rangle/\langle n_e \rangle^{5/3} $) in $\kent$ and $t_{\rm TI}/t_{\rm ff}$~({\em right}) as a function of radius for the group run with feedback heating~($\epsilon_{\rm fb}=7\times10^{-5}$; run G50-fb-7em5 in Table \ref{tab:tab3}). Solid black line represents the initial profile. The other lines are the profiles averaged over a 1~Gyr timescale at different times. Cooling and feedback affect a larger radius in lower mass halos because of the shorter cooling time ~(e.g., compare this Figure~with Figs. \ref{fig:ent_C10} \& \ref{fig:ent_C10_fb}).
\label{fig:ent_G50_fb}}
\end{figure*}

Figure~\ref{fig:ent_G50_fb} shows the entropy and $t_{\rm TI}/t_{\rm ff}$ profiles for the group run with $\epsilon_{\rm fb} = 7 \times 10^{-5}$~(G50-fb-7em5). The qualitative behavior is very similar to the cluster simulations with intermediate feedback. There is an initial spike in accretion because heating is small at early times. After an early episode of filament condensation and heating the critical ratio $t_{\rm TI}/t_{\rm ff}$ becomes $>10$ and there is a period of $\approx 1.5$~Gyr~(from 1.5 to 3~Gyr) without multi-phase gas. There is a second episode of filaments at $\approx 3.5$~Gyr. One important difference from the cluster runs is that the relative size of the core affected by cooling and feedback is much larger for groups than for clusters~(see also Fig.~\ref{fig:nTvsr_sims}).

\subsection{Comparison to Idealized Heating Simulations}
\label{sec:comp}
\begin{figure}
\centering
\includegraphics[scale=0.42]{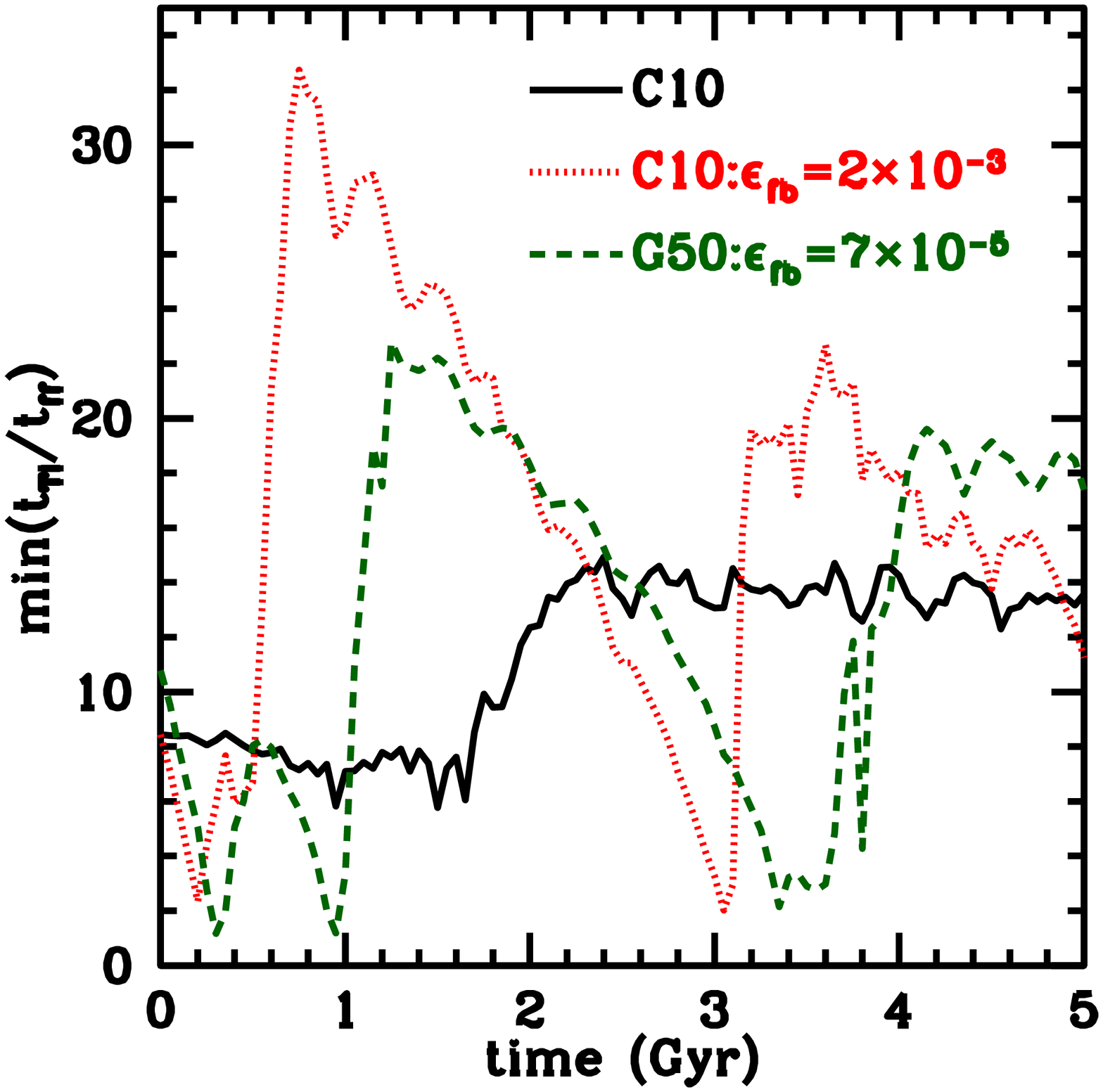}
\caption{The minimum of the ratio $t_{\rm TI}/t_{\rm ff}$ for the fiducial run and the feedback heating simulations sampled every 50~Myr as a function of time~($t_{\rm TI}/t_{\rm ff}$ is evaluated from angle averaged radial profiles of plasma with $T>$ 0.1~keV; the results do not depend sensitively on the cutoff temperature as long as it is $< 1/3$ the ambient temperature). Filaments condense out of the hot phase whenever $t_{\rm TI}/t_{\rm ff} \lesssim 10$ for both idealized and feedback heating runs. This can be seen by comparing this plot with the mass accretion rate as a function of time~(Figs. \ref{fig:mdot}, \ref{fig:mdot_C10_fb}, \& \ref{fig:mdot_G50_fb}). The spike in the mass accretion rate corresponds to infalling cold filaments; it is preceded by a drop in min$(t_{\rm TI}/t_{\rm ff})$ to less than the critical value~($\approx 10$) and followed by an increase in min$(t_{\rm TI}/t_{\rm ff})$ as the ambient plasma is heated and cold matter drops out of the hot phase.
\label{fig:trat_min}}
\end{figure}

Figure~\ref{fig:trat_min} illustrates the feedback regulation process under
each of our heating prescriptions.  Here we plot the minimum of the
angle-averaged $t_{\rm TI}/t_{\rm ff}$ as a function of time, taking only the
X-ray emitting plasma into account; this quantity determines whether the ICM
develops multi-phase structure, and its temporal evolution depends
qualitatively on the heating mechanism.  With our `idealized' heating prescription,
the heating responds both locally and instantaneously to cooling and it keeps
the ICM very close to the threshold for filament formation, $t_{\rm TI}/t_{\rm ff} \sim 10$.  Our `feedback' heating
prescription is not so tailored and it, by contrast, only balances cooling in an average
sense.  Consequently, the departures from the critical value of $t_{\rm TI}/t_{\rm ff}$ are much larger
in our feedback simulations than in our idealized simulations; this leads to
the episodic condensation of cold filaments discussed in
section~\ref{sec:fb_cl}.

Though simulations with our idealized heating prescription are always in
global thermal equilibrium, our feedback simulations are initialized very far
from equilibrium.  As discussed above~(section~\ref{sec:fb_cl}), this leads to a
period of rapid, cold-mode accretion~(around 0.2 Gyr in
Figs.~\ref{fig:mdot_C10_fb}~\&~\ref{fig:mdot_G50_fb}), followed by a strong
feedback event.\footnote{The delay between condensation of cold filaments and
large energy injection due to feedback is of order the free-fall time of
filaments from the radius at which they form.  Also notice that the secular
cooling time in the state without filaments is longer for groups than for
clusters because of a longer cooling/free-fall time in a bigger core~(see
Fig. \ref{fig:nTvsr_sims}).}  This is seen as a sudden increase in min$(t_{\rm
TI}/t_{\rm ff})$, after which heating is suppressed because mass accretion
occurs only through the much slower hot-mode~(see Figs.~\ref{fig:mdot_C10_fb}
and~\ref{fig:mdot_G50_fb}).  As the core slowly cools, the ratio $t_{\rm
TI}/t_{\rm ff}$ again falls below the threshold of $\sim 10$, leading again to
a period of cold-mode accretion and a strong feedback event.  Eventually the
feedback simulations settle in rough thermal balance similar to the idealized
heating runs.

There are several additional similarities between our idealized heating
simulations and the feedback simulations.  The luminosity at sub-virial
temperatures~(as in Fig.~\ref{fig:lum_vs_T}) for feedback simulations looks
very similar to the idealized heating simulations.  In both cases cooler gas
exists only when the ratio $t_{\rm TI}/t_{\rm ff}\lesssim 10$, and cooling to
sub-virial temperatures is suppressed by a factor of $\sim 100$ as compared to
the cooling-flow simulations, in agreement with observations of cool-core
clusters~(e.g., \citealt{pet06}).  Figure \ref{fig:time_vs_mdot} shows that
the time spent accreting at a given $\dot{M}$ is very similar for our
idealized heating simulations and our feedback simulations, for both groups
and clusters.  The fact that the more realistic feedback heating runs are much
closer to our idealized heating runs than to a cooling flow and consistency
with several cluster observations strongly suggest that our models with
idealized/feedback heating capture the key thermal physics of cool-core
clusters.

\section{Additional Astrophysical Implications}

Our work on local thermal instability in global  thermal equilibrium is relevant to several other puzzles related to the hot gas in massive halos. Here we discuss three applications in detail.
 
 \subsection{AGN Fueling: Hot or Cold Gas?}

For massive elliptical galaxies, \citet{all06} argued that the
observed jet power~(calculated from estimating the energy and
inflation timescale for X-ray cavities/radio-bubbles) is well
correlated with the estimated Bondi accretion rate of the hot ICM~(see
\citealt{nar11} for models of this physics). The estimate of the Bondi
accretion rate is quite uncertain because it requires extrapolating
the density and temperature from large radii to the radius of
influence of the supermassive black hole. Nonetheless, the possible
correlation of the Bondi accretion rate and the jet power suggests
that the supermassive black hole may be fed by the hot ICM.  In our
calculations, we indeed find that systems spend most of their time
accreting primarily via the hot phase (Fig. \ref{fig:time_vs_mdot}).

For several reasons, however, it is not clear whether the connection
between the Bondi accretion rate and the true AGN accretion rate is
that robust.  The Bondi accretion rate is $\propto K^{-3/2}$ (where $K$ is the entropy at the Bondi radius).  Thus
lower entropy implies a larger Bondi accretion rate.  However, lower
entropy also implies smaller $t_{\rm cool}/t_{\rm ff}$ (see Eqs. \ref{eq:tTI}-\ref{eq:tff} for the definition of different important timescales) and thus a
higher accretion rate via cold gas formed by local TI.  \citet{cav08}
showed that clusters with the central entropy $\lesssim 30 \kent$~(or equivalently, $t_{\rm TI}/t_{\rm ff} \lesssim 10$; see Eq. \ref{eq:ent_th}) show a large H$\alpha$ luminosity and a large radio luminosity~(a proxy for jet power), suggesting a
connection between cold gas, AGN feedback, and the state of the hot gas. Indeed, our TI model predicts that the mass accretion rate, and 
hence feedback, increases dramatically whenever $t_{\rm cool}/t_{\rm ff} \lesssim 10$. 
By contrast, the dependence of AGN accretion and feedback on the thermal state of the ICM is not expected to be as abrupt if accretion proceeds primarily via hot gas.

For the specific systems in \citet{all06}, the central gas densities are
$\gtrsim 0.1$~cm$^{-3}$. Thus, these systems are in fact prone to
forming cold filaments at small radii according to our $t_{\rm
  cool}/t_{\rm ff}$ criterion. 
Future observations with density/temperature measurements close to the
black hole's sphere of influence are required to ascertain if the
Bondi scaling of the jet power is robust, or if it is by product of
the correlation between entropy~(or, better, $t_{\rm cool}/t_{\rm
  ff}$) and the presence of cold gas via local TI. 
 
 \subsection{$L_X-T_X$ Relation}
 
 The observed relationship between the X-ray luminosity and the X-ray
 emissivity-weighted temperature of groups is steeper~($L_X \propto
 T_X^3$) than the self-similar scaling prediction~($L_X \propto
 T_X^2$; e.g., \citealt{evr91}; \citealt{bry00} and references
 therein).  Here we argue that the non-self similarity introduced by
 the requirement $t_{\rm cool}/t_{\rm ff} \gtrsim 10$ can
 qualitatively account for these observations.

 The total X-ray luminosity produced by a hot halo is
 $$
 L_X = \int_0^{r_{200}} n_e n_i \Lambda d^3r,
 $$
 where $r_{200}$ is the virial radius.
 For simplicity, assume that the plasma in groups and clusters emits via
 thermal free-free emission, such that $\Lambda \propto T^{1/2}$, and
 that the temperature profile of a given halo is
 isothermal.\footnote{These assumptions are not strictly correct but
   do not qualitatively affect our argument.}  In this case
 $$
 L_X = 4 \pi r_s^3 \Lambda_0~(T/T_0)^{1/2} \int_0^{c} n(x)^2 x^2 dx,
 $$ 
 where $\Lambda_0 \sim 10^{-23}$~erg~cm$^3$~s$^{-1}$ is the emissivity
 at $T_0 \approx 10^7$~K, $r_s=r_{200}/c$ is the scale radius, $c$ is the
 concentration parameter, and $x=r/r_s$.  The integrated X-ray
 luminosity is dominated by the core radius $r_c$, at which the
 density profile transitions from being relatively shallow to steep;
 the X-ray luminosity then scales as $L_X \propto r_c^3 T^{1/2}
 n_c^2$, where $n_c$ is the gas number density in the core.
 
 For self-similar evolution of baryons following the dark matter~(\citealt{kai86})
 the gas number density is independent of the halo mass; thus, $L_X
 \propto r_{200}^3 T^{1/2} \propto T^2$, since $r_{200} \propto M^{1/3}$ and
 $T \propto M/r_{200} \propto M^{2/3}$. However, the assumption of a
 constant core density irrespective of the halo mass is not consistent
 with our results, especially for low halo masses where cooling is
 very important. In our calculations, the X-ray emitting plasma in the
 core of groups and clusters has a lower density for a lower mass halo
~(e.g., see Fig.~\ref{fig:nTvsr_sims}). This is because $t_{\rm
   cool}/t_{\rm ff} \gtrsim 10$ irrespective of the halo mass implies
 that $t_{\rm cool}/t_{\rm ff} \propto T^{1/2}/n_c \approx$ constant
 for all halo masses, i.e., $n_c \propto T^{1/2}$.
 Assuming for simplicity that the core radius follows the self-similar
 scaling $r_c \propto M^{1/3}$~(which is similar to what is observed
 but not likely to be correct in detail) we find that
 $$ L_X \propto n_c^2 T^{1/2} r_c^3 \propto T^3,$$ or, more precisely, that the
 maximum X-ray luminosity at a given halo mass is $\propto T^3$.  This
 result is tantalizingly similar to that observed, but a more detailed
 one-dimensional model~(e.g., \citealt{voi02}) is required to
 quantitatively compare our predictions with observations. Such a
 model will be presented in the near future~(Sharma et al., in prep.).

 \subsection{HVCs, TI, \& Density of the Galactic Halo}
  
 Much of the gas falling into the Galactic halo is expected to
 virialize because the halo mass $\approx 10^{12} \msun$~(e.g.,
 \citealt{xue08}) is larger than the critical halo mass required to
 form a virial shock \citep[e.g.,][]{ker09}.  Local TI, however, can
 generate cold gas if the halo density is high enough.  Indeed, some
 fraction of the high velocity clouds~(HVCs) of neutral Hydrogen
 observed in the Galactic halo may result from local TI~(e.g.,
 \citealt{mal04}).

 Figure~\ref{fig:nTvsr_sims} shows that for our models of hot halos, a
 smaller halo mass implies that the size of the density core increases
~(relative to the virial radius) and the core density decreases.  We
 can generalize this result
 to Galactic halo masses~(with temperatures $\sim {\rm few} \times 10^6$~K) 
 in which free-free emission is sub-dominant relative to cooling by
 lines, resulting in an even shorter cooling time. Our one-dimensional
 models of hot halos based on a $t_{\rm TI}/t_{\rm ff}$ threshold
~(Sharma et al., in prep.) show that the core radius for a
 Galactic-mass halo is close to the virial radius~($\approx 100$~kpc)
 and the core density is $n_e \sim 10^{-4}$~cm$^{-3}$. This is
 consistent with the currently poor constraints on the properties of
 the Galactic halo~(e.g., \citealt{and10}).
  
 Given these parameters, local TI can produce cold gas~(e.g., HVCs in
 the Galactic context) at $\sim 100$~kpc, where the ratio $t_{\rm
   cool}/t_{\rm ff}$ is smallest and close to our threshold for
 forming nonlinear structure. Although the gas condenses at large
 radii it falls inward because of gravity and may be observed at much
 smaller radii.  Observations in M31, where the HVC distances are
 easier to measure, have found HVCs as far out as 50~kpc~(and perhaps
 even farther; \citealt{thi04}).  If the HVCs are in pressure balance with the 
 hot halo their emissivity is expected to be smaller at large radii where the pressures are lower. 
 The distribution of cold gas as a
 function of radius is difficult to predict with certainty because of
 uncertainties in the destruction of the cold filaments by conduction,
 ram pressure stripping, etc.
     
 \citet{bin09} argued that the Galactic HVCs cannot form due to local
 TI.  As we discuss in the Appendix, \citet{bin09}'s arguments are
 appropriate for homogeneous cooling-flows with very small amplitude
 density perturbations.  For halos in approximate global thermal
 equilibrium, however, TI can generate multi-phase structure~(and thus
 potentially HVCs) if $t_{\rm TI}/t_{\rm ff} \lesssim 10$.  Even cooling-flows can lead to multi-phase gas if {\em both} $t_{\rm TI}/t_{\rm ff}$ is small enough and there are large amplitude initial density perturbations (see Appendix \ref{app} for details).  We discuss
 the differences between our results and those of \citet{bin09} in
 more detail in \citetalias{mcc11}.

 Although cooling-flow models can, under some circumstances, produce multi-phase
 structure analogous to HVCs, we have shown that the presence of
 approximate global thermal equilibrium in hot halos makes the
 formation of multi-phase structure far more generic, e.g., even with very small density perturbations.  This is true
 even if there are substantial~(order unity) fluctuations in heating
 and cooling about the global equilibrium~(see \citetalias{mcc11}
 and section \ref{sec:robust}).

\section{Summary}

The primary source of baryons in galaxies is the smooth accretion of
intergalactic gas into dark matter halos~(as opposed to mergers; this is particularly true at the current
redshift; e.g., see \citealt{dek09,ker09}). For halo masses $\lesssim
10^{11.5} \msun$~(such that the cooling time is $\lesssim$ the
free-fall time at the virial radius; e.g., \citealt{ree77}) the
accreted baryons remain cold inside the virial radius of the
halo. For more massive halos, the cooling time at the virial radius is
longer than the free-fall time and the virial shock creates
pressure-supported hot plasma that fills the dark matter halo.  The
focus of this paper is on this latter mode of accretion, relevant for
clusters, groups, and massive galaxies.  Although the cooling time is
longer than the free-fall time at the virial radius, the cooling time
can be significantly shorter than the Hubble-time~(and in principle
even shorter than the free-fall time) at small radii if the gas
density profile were to approximately follow the dark matter halo
density profile. Thus, cooling has a profound impact on the hot ICM in
massive halos.

If the cooling time is shorter than the `age' of a halo~(which is
approximately the time since the last major merger in a realistic
cosmological context), a massive cooling-flow will develop absent
additional heating. Thus, the observational absence of cooling-flows
(e.g., \citealt{pet06}) and the lack of star formation rates of the order of the
expected cooling rates~(e.g., \citealt{raf08}) imply that there is
an additional heat source roughly balancing cooling in the hot halos
of groups and clusters. Other proposals such as a uniform preheating
(e.g., \citealt{toz01}) may be important for systems with high central
entropies and long central cooling times, but cannot explain the
presence of long-lived cool-core clusters~(e.g., \citealt{mcc08}).

While there are many potential sources of energy that can maintain
approximate thermal equilibrium,
the particularly attractive feature of heating by star formation
and/or an AGN is that it can be {\em globally} thermally stable~(e.g.,
see \citealt{guo08} for a specific realization of this): the
increasing feedback due to the cooling and inflow of gas provides a
thermostat which prevents the ICM core from run-away cooling.  In this
paper we have addressed a previously under-appreciated aspect of this
scenario: even if heating can balance cooling in a globally stable
manner, and even if the heating can be distributed over the full
volume of the ICM~(a difficult problem;
\citealt{ver06,sca08,dub10,gas11}), the cooling of the ICM is very
likely to generate {\em local} thermal instability~(TI).

We have addressed a number of key science questions associated with
local TI in hot halos.  The first is the condition under which local
TI in a globally stable ICM produces a multi-phase medium with cold gas
co-spatial with the hot ICM.  We find that the criterion for TI to
produce multi-phase gas in global spherical numerical simulations is that the
thermal instability growth time $t_{\rm TI}$ satisfy $t_{\rm TI}
\lesssim 10~t_{\rm ff}$~(see Fig.~\ref{fig:trat_min_1} \& the left panel of Fig.~\ref{fig:mdot}), where $t_{\rm ff}$ is the local free-fall
time~(note that this criterion is derived from global simulations but
it applies at each location within a hot halo; see Eqs. \ref{eq:tTI}-\ref{eq:tff} for the definition of these timescales).  When $t_{\rm TI}
\gtrsim 10 \, t_{\rm ff}$, the TI leads to density and temperature
fluctuations but they do not become nonlinear and so multi-phase gas
cannot generically be produced from arbitrary initial conditions.

In \citetalias{mcc11}, we presented a closely related study of local
TI in globally stable systems using analytic theory and local
Cartesian simulations.  A more comprehensive discussion of the key
physics is given there.  In the local simulations, the criterion for
TI to produce multi-phase gas is that $t_{\rm TI} \lesssim t_{\rm ff}$
(rather than $t_{\rm TI} \lesssim 10~t_{\rm ff}$ as we find
here). Thus it is easier for cold gas to condense out of the ICM in
spherical geometry than in Cartesian geometry.  Our interpretation is
that this difference is because in spherical geometry a slightly
overdense fluid element is compressed as it falls; this increases the
density and reduces the cooling time, aiding the development of the
TI.\footnote{In addition to spherical compression, the background entropy 
stratification~(e.g., \citetalias{mcc11}) and the size distribution of cool `blobs'~(e.g., \citealt{piz05}) 
may also affect the formation of cold filaments.}  
Our criterion for the production of multi-phase gas via TI does not depend
sensitively on how we implement the uncertain heating of the ICM.  We
have studied two different models: an `idealized' heating model in which the
heating in a given radial shell is equal to the average cooling within
that shell~(Eq. \ref{eq:heat}) and a `feedback' heating model in which the heating is set
by the mass accretion rate at the smallest radii~(Eq. \ref{eq:heat_fb}).  Both
heating models produce similar results overall and quantitatively
similar criterion for the production of multi-phase gas via TI.

Figure~12 in \citetalias{mcc11} shows the observed radial profiles of
$t_{\rm cool}/t_{\rm ff}$ for a number of clusters, some of which show
spatially extended H$\alpha$ emission and some of which do not.  The
transition between clusters with and without this observational
signature of intracluster cold gas occurs at $t_{\rm cool}/t_{\rm ff}
\sim 5-10$, consistent with our results on the origin of multi-phase
gas via TI in the ICM.

The formation of multi-phase gas via TI in hot halos makes a number of
other predictions about the observed properties of the ICM that are in
reasonable agreement with observations.  We highlight several of the
key ones here.  First, the {\em net} cooling rate to low temperatures
(and the mass accretion rate at small radii) for models in global
thermal balance is $10^{2-3}$ times smaller than for cooling-flow
models, consistent with observations~(e.g.,
\citealt{pet03,ode08}). The net cooling rate is the largest when
$t_{\rm TI}/t_{\rm ff} \lesssim 10$ and cold gas is condensing out of
the hot phase.  Even for models with $t_{\rm TI} \sim 10~t_{\rm ff}$,
however, which is typical of cool-core clusters, the net cooling rate
is a factor of $\sim 100$ smaller than the cooling-flow value for that
cluster~(e.g., Fig.~\ref{fig:lum_vs_T}).  This is a nontrivial result
that is a consequence of {\it both} global thermal balance and the
physics of local TI in gravitationally stratified plasmas.  A
corollary of this result is that in a given halo the observed X-ray
emission is produced almost entirely by gas within a factor of a few of
the virial temperature~(Fig.~\ref{fig:lum_vs_T_Mh}).  This is consistent
with observational results from X-ray spectroscopy
\citep[e.g.,][]{pet03,san10}.  Previous authors have searched for
heating mechanisms that can self-adjust to reproduce this narrow range
in temperatures in a given halo~(e.g., \citealt{kim05}).  Our results
suggest that these observations are instead a natural consequence of
the approximate isothermality of halo potentials, together with the
physics of local TI for $t_{\rm TI} \gtrsim 5-10~t_{\rm ff}$, which
results in only a very small fraction of the gas at any time cooling
to sub-virial temperatures.

A second key consequence of the formation of multi-phase gas via TI is
related to how the global thermal balance of hot halos is maintained
in the first place.  Figure~\ref{fig:mdot} shows that the accretion
rate at small radii increases dramatically, by up to $\sim 3$ orders
of magnitude, when $t_{\rm TI} \lesssim 10~t_{\rm ff}$ and cold gas
condenses out of the hot ICM.  In any model in which the heating of
the hot halo is related to this inflow rate~(e.g., heating via an AGN
or star formation), the heating of the ICM will increase dramatically
if $t_{\rm TI} \lesssim 10~t_{\rm ff}$.  This will in turn drive
$t_{\rm TI}/t_{\rm ff} \gtrsim 10$.  As a result hot
halos will tend to self-adjust such that $\min(t_{\rm TI}/t_{\rm ff})
\gtrsim 10$.  Both our `idealized' heating and `feedback' heating
models demonstrate this explicitly~(Figs.~\ref{fig:trat_min_1} \&
\ref{fig:trat_min}).  We suggest that this physics is responsible for
the significant fraction of clusters that have central cooling times
$\sim 3 \times 10^{8}$ yrs~(e.g., \citealt{cav09}), which corresponds
to a minimum value of $t_{\rm cool} \sim 10~t_{\rm ff}$ in cluster
cores~(see Fig.~12 of \citetalias{mcc11}).

The net heating efficiency required to maintain approximate global
thermal equilibrium in clusters is $\sim 1-3 \times 10^{-3}$~(we define the
efficiency as the heating integrated over the region with a cooling
time $<$ 5~Gyr divided by the rest mass energy accreted through our
inner boundary at $\lesssim 1$~kpc).  This efficiency can plausibly be
provided by a central AGN.  For lower mass halos, the efficiency
required for thermal equilibrium is lower.  Extrapolating to halo
masses $\sim 10^{12} \msun$, an efficiency of $\lesssim 10^{-5}$ would be 
more than sufficient for global thermal equilibrium with min($t_{\rm TI}/t_{\rm
  ff}$) $\gtrsim 10$ (Sharma et al., in prep.).  This is comparable to the efficiency of energy
injection by core-collapse supernovae, implying that supernovae
venting from galactic disks may be sufficient to maintain global
thermal equilibrium in halos like that of the Milky Way.

In this paper we have focused on how cold gas-rich filaments condense out of a hot halo; we have not, however, addressed important questions about the structure and stability of the cold gas. High-resolution numerical simulations with magnetic fields and anisotropic thermal conduction are required to study the survival of the filaments as they fall through the hot gas. Arguments based on mixing due to Kelvin-Helmholz instability suggest that once a large-enough~($\gtrsim 100$ pc) blob cools out of the hot phase it survives its passage through the hot medium~(e.g., see \citealt{mur93,piz05}). Magnetic fields further stabilize the cold blobs against mixing so we expect the cold filaments to survive as they fall toward the center, as is observed in many cases.

In section 5 we briefly summarized the implications of our results for other problems related to the astrophysics of hot halos, specifically the relative importance of hot and cold gas for AGN fueling in clusters (5.1), the non self-similarity of the relationship between X-ray luminosity and temperature for groups and clusters (5.2), and the possible origin of high velocity clouds of neutral Hydrogen in the local group via TI in hot halos (5.3).  These applications will all be studied in more detail in future work.

\section*{Acknowledgments}

We thank Du\v{s}an Kere\v{s}, Leo Blitz, and Arif Babul for useful discussions. Support for P. S. was provided by NASA through Chandra Postdoctoral Fellowship grant PF8-90054 awarded by the Chandra X-Ray Center, which is operated by the Smithsonian Astrophysical Observatory for NASA under contract NAS8-03060. M. M., I. P. and E. Q. were supported in part by NASA Grant NNX10AC95G, NSF-DOE Grant PHY-0812811, and the David and Lucile Packard Foundation. Computing time was provided by the National Science Foundation through the Teragrid resources located at Purdue University and NCSA. Some of the runs were carried out on {\em Henyey}, the theoretical astrophysics computing cluster at UC Berkeley supported by NSF AST Grant 0905801.  We are also thankful for the hospitality of the Kavli Institute for Theoretical Physics~(KITP) at UC Santa Barbara, where part of this work was performed.

\onecolumn 

\begin{table}
\caption{Runs with Idealized Heating \label{tab:tab1}}
\resizebox{\textwidth}{!}{%
\begin{tabular}{cccccccccccc}
\hline
\hline
Label$^\dag$ & $K_0^\ast$ & $K_{100}^\ast$ & $\alpha^\ast$ & $t_{\rm TI}/t_{\rm ff}^\ddag$ & $\dot{M}_{>0.1}^{\dag\dag}$ & $\dot{M}_{0.01-0.1}^{\dag\dag}$ & $\dot{M}_{<0.01}^{\dag\dag}$ & $M_{>0.1}^{\dag\dag}$ & $M_{0.01-0.1}^{\dag\dag}$ & $M_{<0.01}^{\dag\dag}$ & $\epsilon^{\ddag\ddag}=Q^+/\dot{M}c^2$  \\
\hline
MC1 & 1.91 &  209.7 & 1.4 &  2.9~(2.3) & 7.3~(70.3) & 0.02~(1) & 12.5~(470.6) & $9.4 \times 10^{12}$~($9 \times 10^{12}$) & $4.3 \times 10^6$~($3.8 \times 10^7$) & $4.3 \times 10^8$~($8.1 \times 10^9$) & $8.2 \times 10^{-4}$ \\ 
MC10 & 19.1 & 209.7 & 1.4 & 32.5~(12.5) & 2.6~(41.4) & 0~(0.5) &  0~(348.3) & $8.8 \times 10^{12}$~($8.9 \times 10^{12}$) & 0~($3.5 \times 10^7$) & 0~($6.7\times 10^9$) & $3.7 \times 10^{-3}$ \\
MC30 & 57.2 & 209.7 & 1.4 & 111.4~(32.5)  & 0.8~(21.1) & 0~(1.2) & 0~(169) & $8 \times 10^{12}$~($8.4 \times 10^9$) & 0~($1.5 \times 10^8$) & 0~($4 \times 10^9$) & $4.3 \times 10^{-3}$  \\ 
MC100 & 190.6 & 285.9 & 0.8 & 477~(128.5) & 0.3~(0.4) & 0~(0) & 0~(0) & $6 \times 10^{12}$~($6.3 \times 10^{12}$) & 0~(0) & 0~(0) &  0 \\ 
MC300 & 571.8 & 152.5 & 0.8 & 1495~(191) & 0.2~(0.5) & 0~(0) & 0~(0) & $5.1 \times 10^{12}$~($5.2 \times 10^{12}$) & 0~(0) & 0~(0) & 0 \\
C1 & 1 &  110 & 1.4 &  1.6~(1.4) & 2.8~(13.7) & 0.03 ~(3.7) & 9.1~(285.6) & $3.5 \times 10^{12}$~($3.4 \times 10^{12}$) & $4.3 \times 10^6~(4.3 \times 10^{8}) $ & $3.5 \times 10^8 $~($5.6 \times 10^9$) & $ 4.5 \times 10^{-4}$ \\ 
C10$^\star$ & 10 & 110 & 1.4 & 20.2~(8.2) & 1.2~(11.4) & 0.003~(3.04) & 2.3~(230.3) & 3.3 $\times 10^{12}$~($3.3 \times 10^{12}$) & $2.6 \times 10^6$~($ 3.6 \times 10^8 $) & $1.3 \times 10^8 $~($4.9 \times 10^9$) & $10^{-3}$ \\
C30 & 30 & 110 & 1.4 &  71~(21.6) & 0.4~(7.7) & 0~(1.2) & 0~(144.6) & 3$\times 10^{12}$~($3.2 \times 10^{12}$) & 0~($2.5 \times 10^8$) & 0~($4 \times 10^9$) & $ 5 \times 10^{-3} $ \\ 
C100 & 100 & 150 & 0.8 &  314.4~(85) & 0.1~(0.3) & 0~(0) & 0~(0) & 2.3$\times 10^{12}$~($2.5 \times 10^{12}$) & 0~(0) & 0~(0) & $5.4 \times 10^{-5}$ \\ 
C300 & 300 & 80 & 0.8 & 1001~(126.4) & 0.1~(0.2) & 0~(0) & 0~(0) & $2 \times 10^{12}$~($2 \times 10^{12}$) & 0~(0) & 0~(0) & 0 \\
MG1 & 0.5 &  57.7 & 1.4 & 0.52~(0.46) & 0.4~(2.7) & 0.02~(0.3) & 16.2~(130.4) & 8.9 $\times 10^{11}$~($8.6 \times 10^{11}$) & $3.9 \times 10^7$~($3.2 \times 10^7$) & $1.3 \times 10^9$~($3.3 \times 10^9$) & $7 \times 10^{-5}$ \\ 
MG10 & 5.3 & 57.7 & 1.4 & 9.9~(3.8) & 0.3~(2.4) & 0.02~(0.2) & 4.7~(113.9) & $8.6 \times 10^{11}$~($8.4 \times 10^{11}$) & $2 \times 10^{7}$~($2.8 \times 10^7$) & $2.91 \times 10^{8}$~($2.9 \times 10^9$) &  $1.8 \times 10^{-4}$ \\
MG30 & 15.7 & 57.7 & 1.4 &  37.1~(10.7) & 0.2~(1.9) & 0~(0.1) & 0~(89.5) & $7.9 \times 10^{11} $~($8.2 \times 10^{11}$) & 0~($2.2 \times 10^7$) &  0~($2.4 \times 10^9$) & $3.2 \times 10^{-3}$ \\ 
MG100 & 52.5 & 78.7 & 0.8 & 169.1~(45.8) & 0.04~(0.6) & 0~(0.02) & 0~(27.4) & $6 \times 10^{11}$~($7.3 \times 10^{11}$) & 0~($1.7 \times 10^7$) & 0~($1.4 \times 10^9$) & $1.2 \times 10^{-3}$ \\ 
MG300 & 157.4  & 42 & 0.8 & 557.4~(68.7)  & 0.02~(0.02) & 0~(0) & 0~(0) & $5.2 \times 10^{11}$~($5.9 \times 10^{11}$) & 0~(0) & 0~(0) & 0 \\
G5 & 1.1 &  23.7 & 1.4 & 1.8~(0.9) & 0.07~(0.5) & 0.004~(0.07) & 8.6~(73.7) & $3.2 \times 10^{11}$~($3.1 \times 10^{11}$) & $2 \times 10^7$~($3.8 \times 10^7$) & $8.8 \times 10^8$~($2.5 \times 10^9$) & $4 \times 10^{-5}$ \\ 
G50 & 10.8 & 23.7 & 1.4 & 39.7~(10.4) & 0.05~(0.4) & 0.003~(0.05) & 2.3~(48.9) & $2.7 \times 10^{11}$~($3 \times 10^{11}$) & $1.2 \times 10^7$~($2.3 \times 10^7$) & $2 \times 10^8$~($1.6 \times 10^9$) & $7.4 \times 10^{-5}$ \\
G75 & 16.2 & 23.7 & 1.4 &  65.8~(16.2) & 0.03~(0.13) & 0~(0.4) & 0~(40.9) & $2.6 \times 10^{11}$~($2.9 \times 10^{11}$) & 0~($1.1 \times 10^9$) & 0~($1.3 \times 10^9$) & $4.7 \times 10^{-3}$ \\ 
G150 & 32.3 & 17.2 & 0.8 &  158.6~(26.4) & 0.01~(0.12) & 0~(0.2) & 0~(41.6) & $2.3 \times 10^{11} $~($2.9 \times 10^{11}$) & 0~($9.7 \times 10^8$) & 0~($1.5 \times 10^9$) & $ 5.3 \times 10^{-3}$ \\ 
G300 &  64.6 & 17.2 & 0.8 & 350.3~(42.3) & 0.006~(0.05) & 0~(0.04) & 0~(3.9) & $2 \times 10^{11}$~($2.3 \times 10^{11}$) & 0~($6 \times 10^7$) & 0~($7.8 \times 10^7$) & $1.5 \times 10^{-4}$ \\ 
\hline
\end{tabular}}

$\star$ {C10 is the fiducial run.} \\
$\ast$ {The entropy $K~(T_{\rm keV}/n_e^{2/3}$) is initialized via Eq. \ref{eq:ini_ent}.}\\
$\dag$ {C1 is the label for a cluster~($M_0=3.8\times 10^{14} \msun$) with $K_0=1\kent$. MC1 stands for a massive cluster with $M_0=10^{15} \msun$ and a core entropy $K_0$~scaled according to self-similarity, $K \propto M_0^{2/3}$, from the cluster model with $K_0=1\kent$. MG stands for a massive group with $M_0=10^{14} \msun$ and G stands for a group with $M_0=3.8\times 10^{13} \msun$. Initial conditions for all halo masses are scaled with $M_0$ according to self-similarity. Self-similarity is broken with time because of cooling and feedback.} \\
$\ddag$ {initial value of the ratio $t_{\rm TI}/t_{\rm ff}$ evaluated at the inner boundary~(value in the parenthesis is the minimum value over all radii). $t_{\rm ff}$ at the inner boundary is $\approx$ 0.014~Gyr for all cases.} \\
${\dag\dag}$ { $\dot{M}_{>0.1}$ represents the average~(over the whole run for 5~Gyr) mass accretion rate~(in $\mpy$) through the inner boundary for plasma with $T>0.1$~ keV, $\dot{M}_{0.01-0.1}$ for 0.01~keV $<T<$ 0.1~keV, and $\dot{M}_{<0.01}$ for $T<0.01$~keV. Values in parentheses correspond to `cooling-flow' models with no external heating. Similarly, $M_{>0.1}$ is the gas mass~(in $\msun$) in the hot phase, and likewise for $M_{0.01-0.1}$ and $M_{<0.01}$; values in  parentheses correspond to cooling-flow models.} \\
$\ddag\ddag$ { $\epsilon$ represents the time-averaged efficiency required to balance cooling: $\epsilon \equiv Q^+/\dot{M}c^2$, where $\dot{M}$ is the time-averaged total mass accretion rate through the inner boundary, and $Q^+=\int n_en_i \Lambda(T) dV$ and the integral is over all radii where the angle-averaged cooling time is $<$~5~Gyr.} 
\end{table}

\begin{table}
\caption{Additional Cluster Runs \label{tab:tab2}}
\resizebox{\textwidth}{!}{%
\begin{tabular}{ccccccccc}
\hline
\hline
{Label$^\dag$} & $N_r \times N_\theta \times N_\phi$ & {$\dot{M}_{>0.1}$} & {$\dot{M}_{0.01-0.1}$} & {$\dot{M}_{<0.01}$} & {$M_{>0.1}$} & {$M_{0.01-0.1}$} & {$M_{<0.01}$} & {$\epsilon=Q^+/\dot{M}c^2$} \\ 
\hline
C10-p1 & $512 \times 256 \times 1$ & 1.1 & 0.01 & 1.7 & $ 3.3 \times 10^{12}$ & $1.7 \times 10^6 $ & $6.2 \times 10^7$ & $1.3 \times 10^{-3}$ \\ 
C10-p10 & $512 \times 256 \times 1$ & 1.7 & 0.001 & 0.24 & $3.3 \times 10^{12}$ & $3.5 \times 10^4$ & $4.8 \times 10^6$ & $2 \times 10^{-3}$ \\
C10-nz1& $ 512 \times 256 \times 1$ & 0.3 & 0.0009 & 0.089 & $4.2 \times 10^{11}$ & $6.1 \times 10^3$ & $1.1 \times 10^6$ & $3.8 \times 10^{-4}$ \\
C10-nz4 & $512 \times 256 \times 1$ & 1.1 & 0.004 & 2.6 & $2.7 \times 10^{12}$ & $2.9 \times 10^6$ & $1.9 \times 10^8$ & $4.1 \times 10^{-4}$ \\
C10-nz8 & $512 \times 256 \times 1$  & 1.3 & 0.02 & 2.5 & $3.1 \times 10^{12}$ & $2.6 \times 10^6$ & $10^8$ &   $7.1 \times 10^{-4}$ \\
C10-nz16 & $512 \times 256 \times 1$ & 1.4 & 0.001 & 0.44 & $3.3 \times 10^{12}$ & $1.3 \times 10^5$ & $1.2 \times 10^7$ & $2 \times 10^{-3}$ \\
C10-3D & $256 \times 128 \times 32 $  & 0.77 & 0.018 & 0.95 & $3.5 \times 10^{12}$ & $9.1 \times 10^5$ & $1.6 \times 10^8 $ & $2.4 \times 10^{-3}$ \\
C10-hlf & $256 \times 128 \times 1$ & 1.3 & 0.004 & 1.1 & $3.5 \times 10^{12}$ & $4.8 \times 10^5 $ & $3.5 \times 10^7$ & $3 \times 10^{-3}$ \\ 
C10 & $512 \times 256 \times 1$ & 1.2 & 0.003  & 2.3 & 3.3 $\times 10^{12}$ & $2.6 \times 10^6$ & $1.3 \times 10^8 $ & $10^{-3}$ \\
C10-dbl & $1024 \times 512 \times 1$ & 1.4 & 0.002 & 0.85 & $2.6 \times 10^{12}$ & $5.4 \times 10^5$ & $2.6 \times 10^7$ & $7 \times 10^{-4}$ \\
C30-p1 & $512 \times 256 \times 1$ & 0.38 & 0 & 0 & $3 \times 10^{12}$ & 0 & 0 & $4.9 \times 10^{-3}$ \\
C30-p10 & $512 \times 256 \times 1$ & 0.58 & 0 & 0 & $2.9 \times 10^{12}$ & 0 & 0 & $2.8 \times 10^{-3}$ \\
C30-3D & $256 \times 128 \times 32 $ & 0.24 & 0 & 0 & 3.2 $\times 10^{12}$ & 0 & 0 & $8.6 \times 10^{-3}$ \\ 
C30-hlf & $256 \times 128 \times 1$ & 0.28 & 0 & 0 & 3.2 $\times 10^{12}$ & 0 & 0 & $7.6 \times 10^{-3}$ \\ 
C30 & $512 \times 256 \times 1$ & 0.4 & 0 & 0 & 3$\times 10^{12}$ & 0  & 0 & $ 5 \times 10^{-3} $ \\ 
C30-dbl & $1024 \times 512 \times 1$ & 0.36 & 0 & 0 & $3 \times 10^{12}$ & 0 & 0 & $1.5 \times 10^{-3}$ \\
\hline
\end{tabular} }

{$\dag$}{C10-p10 is a cluster run with random perturbations $\approx 10$~(drawn from a uniform distribution) in the heating rate on top of the balance of cooling and heating in each shell; the random perturbations are changed every 100~Myr. C10-nz4 corresponds to a run where the balance of heating and cooling is applied over 4 logarithmically spaced radial zones~(i.e., $H/r = 1/4$, where $H$ is the scale over which heating is applied uniformly). C10-3D is a three-dimensional run with the same parameters as the fiducial run  C10. C10-dbl~(C10-hlf) has double~(half) the fiducial resolution. All remaining quantities are defined in the same way as Table \ref{tab:tab1}.} \\
\end{table}

\begin{table}
\caption{Cluster and Group Runs with Feedback Heating \label{tab:tab3}}
\resizebox{\textwidth}{!}{%
\begin{tabular}{ccccccccc}
\hline
\hline
{Label$^\dag$} & $\epsilon_{\rm fb}$ & {$\dot{M}_{>0.1}$} & {$\dot{M}_{0.01-0.1}$} & {$\dot{M}_{<0.01}$} & {$M_{>0.1}$} & {$M_{0.01-0.1}$} & {$M_{<0.01}$} & {$\epsilon=Q^+/\dot{M}c^2$} \\
\hline
C10-fb-7em3 & $7\times10^{-3}$ & 0.15 & 0.0017 & 0.6 & $2.84 \times 10^{12}$ & $8.28 \times 10^4$ & $ 1.07 \times 10^7$ & $1.4 \times 10^{-3}$ \\
C10-fb-4em3  & $4\times10^{-3}$ & 0.46 & 0.006 & 2.74 & $2.74 \times 10^{12}$ & $2.02 \times 10^6$ & $1.3 \times 10^8$ & $5.8 \times 10^{-4}$ \\ 
C10-fb-2em3  & $2\times10^{-3}$ & 0.77 & 0.0055 & 2.34 & $3.18 \times 10^{12}$ & $1.17 \times 10^6$ & $1.05 \times 10^8$ & $7.2 \times 10^{-4}$  \\
C10-fb-7em4 & $7\times10^{-4}$ & 1.67 & 0.019 & 4.9 & $3.22 \times 10^{12}$ & $3.5 \times 10^6$ & $2 \times 10^8$ &  $3.7 \times 10^{-4}$\\
G50-fb-4em3 & $4\times10^{-3}$ & 0.022 & 0.00042 & 0.079 & $1.8 \times 10^{11}$ & $1.88 \times 10^5$ & $5.79 \times 10^6$ & $2.9 \times 10^{-4}$ \\
G50-fb-2em3 & $2\times10^{-3}$ & 0.029 & 0.00041 & 0.19 & $1.76 \times 10^{11}$ & $1.25 \times 10^5$ & $7.24 \times 10^6$ & $1.1 \times 10^{-4}$ \\
G50-fb-7em4 & $7\times10^{-4}$ & 0.026 & 0.00046 & 0.22 & $2.54 \times 10^{11}$ & $4.3 \times 10^5$ & $1.09 \times 10^7$ & $ 5.3 \times 10^{-4}$ \\
G50-fb-7em5 &  $7\times10^{-5}$ & 0.051 & 0.0058 & 3.23 & $2.99 \times 10^{11}$ & $7.83 \times 10^6$ & $2.18 \times 10^8$ & $7.5 \times 10^{-5}$ \\
\hline
\end{tabular}}

{$\dag$}{C10-fb-7em3 stands for a run with the same initial profile as C10~(the fiducial run) but using a feedback heating model with $\epsilon_{\rm fb} = 7\times10^{-3}$~(see Eq. \ref{eq:heat_fb}). The group runs~(with the same initial conditions as G50) with feedback heating are also included.  All remaining quantities are defined in the same way as Table \ref{tab:tab1}.}
\end{table}

\twocolumn

\appendix
\section{Inhomogeneous Cooling Flows}
\label{app}

\begin{figure*}
\centering
\includegraphics[scale=0.31]{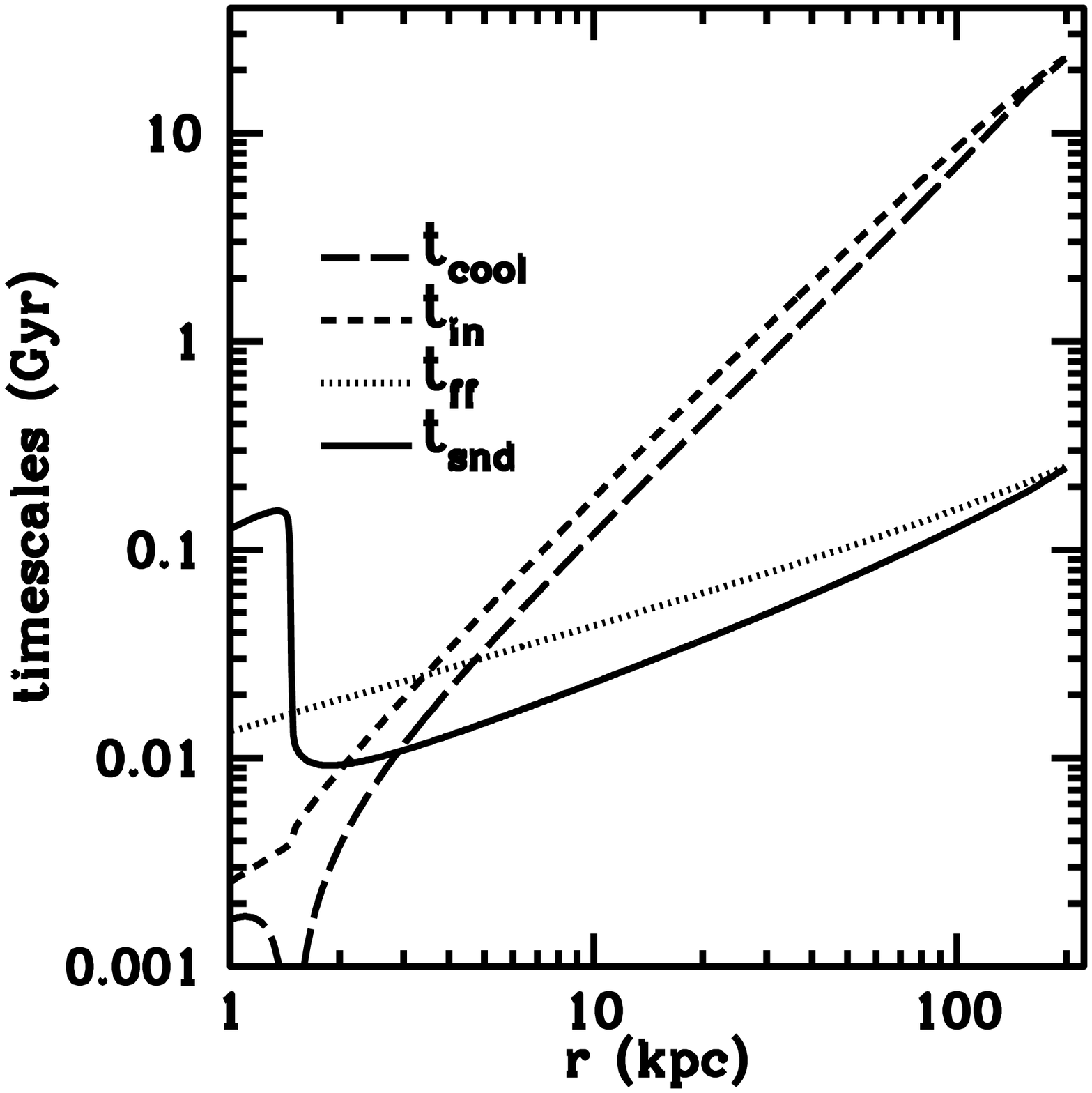}\includegraphics[scale=0.31]{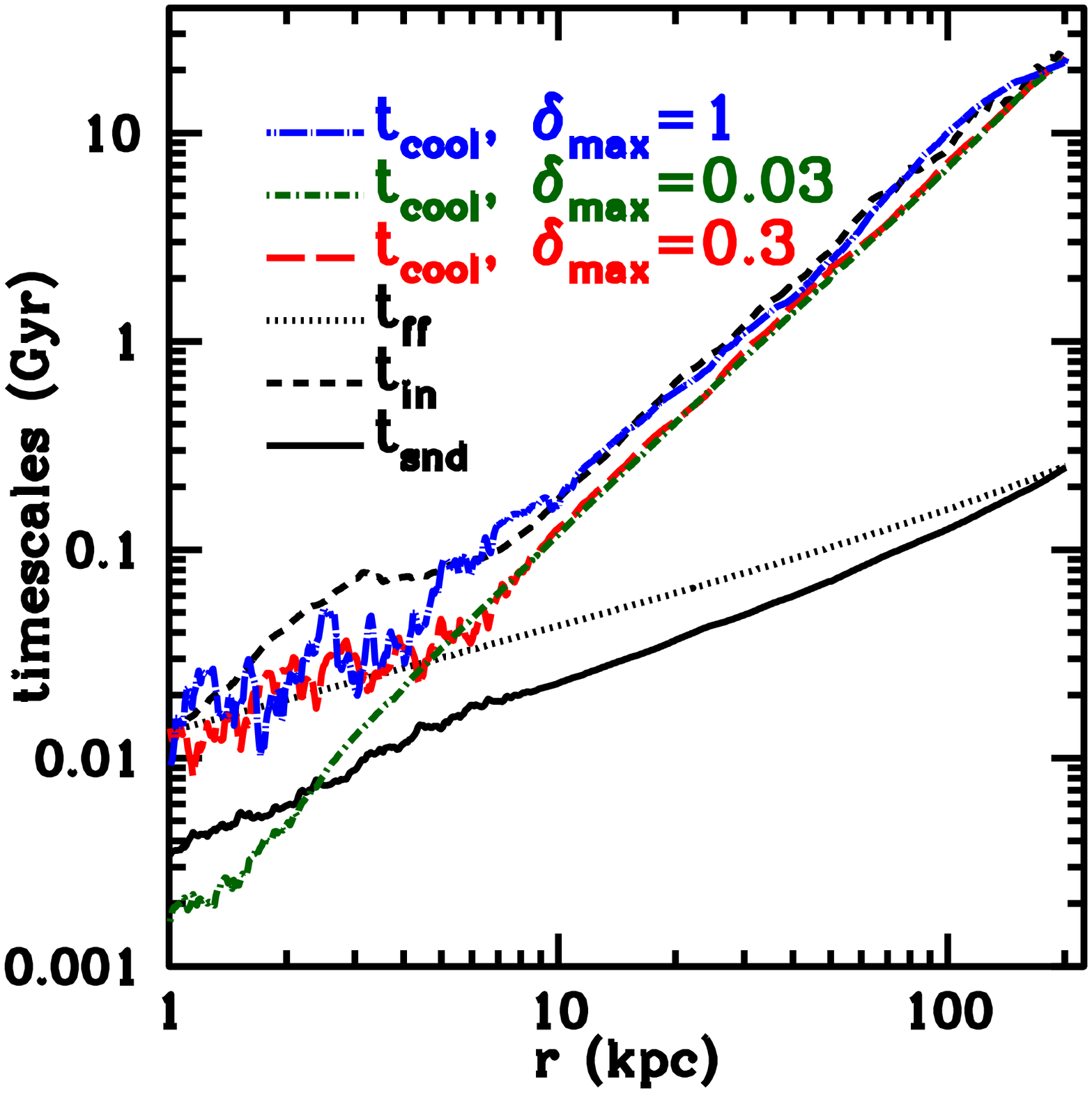}
\caption{Key timescales in a steady homogeneous cooling-flow~({\em
    left} panel) and a quasi-steady multi-phase cooling-flow with large
  density perturbations~({\em right} panel; see
  eqs. \ref{eq:mass_cf}-\ref{eq:en_cf} for the precise definition of
  the different timescales). Only the plasma in the hot~($T>0.1$~keV)
  phase has been used to calculate the relevant timescales for the
  multi-phase cooling-flow simulation. In the right panel the sound
  crossing time~($t_{\rm snd}$) and the infall time~($t_{\rm in}$) are
  only shown for our fiducial density perturbations~($\delta_{\rm max}=0.3$).
  \label{fig:cf_vs_r}}
\end{figure*}

In this Appendix we discuss the nature of hot halos in the absence of
external heating---in this case the halo plasma settles into a
cooling-flow.  We consider both steady homogeneous cooling-flows and
cooling-flows with large initial density perturbations that can
develop multi-phase structure.  Although observations imply that hot,
virialized halos are in approximate thermal equilibrium, the results
in this Appendix are useful for comparison to other results in the
literature and for comparison to the simulations with heating
described in the main text.

\subsection{Steady Homogeneous Cooling Flows}

In a steady~($\partial/\partial t = 0$) cooling-flow the equations of conservation of mass, momentum, and energy can be expressed as
\ba
\label{eq:mass_cf}
&& 4 \pi r^2 \rho v = \dot{M}, \\
\label{eq:mom_cf}
 && -\frac{1}{t_{\rm in}^2} \frac{d\ln v}{d \ln r} = \frac{1}{t_{\rm snd}^2} \frac{d \ln p}{d \ln r} + \frac{2}{t_{\rm ff}^2}, \\
\label{eq:en_cf}
&& \frac{1}{t_{\rm in}} \frac{d \ln \left ( p/\rho^\gamma \right )}{d
  \ln r} = \frac{1}{t_{\rm cool}}, \ea where $v~(> 0)$ is the radial
infall velocity, $\dot{M}$ is the steady mass inflow rate, $t_{\rm in}
\equiv r/v$ is the infall time, $t_{\rm snd} \equiv r/(p/\rho)^{1/2}$
is the isothermal sound-crossing time, $t_{\rm ff}$ is the free-fall
time~(Eq. \ref{eq:tff}), and $t_{\rm cool}$ is the cooling time
(Eq. \ref{eq:tcool}). We use Eqs. \ref{eq:mass_cf}-\ref{eq:en_cf} to
interpret our cooling-flow results discussed shortly.

For our homogeneous cooling-flow simulations we start with an initial
hydrostatic profile identical to the fiducial simulation described in
the main text~(C10), but without density/temperature perturbations. As
in our simulations with heating, we use cosmologically realistic
inflow boundary conditions.  Since the cooling time is the shortest at
the center~(but $t_{\rm cool} \gg t_{\rm snd} \sim t_{\rm ff}$ in the
initial state), gas starts to cool and flow in subsonically~(eq
\ref{eq:en_cf}).  The left panel of Figure~\ref{fig:cf_vs_r} shows the
key timescales in the resulting steady cooling-flow.  As the gas cools
and flows in to small radii, the cooling time becomes less than the
free-fall time at $r \lesssim 5$~kpc.  The gas then abruptly cools to
low temperatures and the inflow becomes supersonic; this corresponds
to the rapid increase~(decrease) in the sound-crossing~(cooling) time
in the left panel of Figure~\ref{fig:cf_vs_r} at $\sim 1.5$~kpc.  For
a given gravitational potential, the radius at which the steady
cooling-flow becomes supersonic is determined primarily by the density
at the outer boundary; for a larger density at large radii the sonic
radius moves out, while for a lower density at large radii, the sonic
radius decreases and the flow may never become supersonic.

\subsection{Multiphase Gas in a Cooling Flow}

Small amplitude density perturbations do not grow significantly via
thermal instability in a cooling-flow because the perturbations are
advected inwards on the same timescale that the thermal instability
would nominally develop \citep{bal89}.  This is a critical difference
between cooling-flows and the plasmas in approximate global thermal
equilibrium considered in the main text.  One implication of this
result is that cooling-flows with small amplitude density
perturbations cannot, in general, develop multi-phase structure.  With
sufficiently large density perturbations, however, a cooling-flow can
produce multi-phase gas~(e.g., \citealt{piz05}). The resulting
structure is very different from the homogeneous cooling-flows
described in the previous section. A rough criterion for whether a
cooling-flow becomes multi-phase can be derived as follows. Multiphase
gas will result if the effective cooling rate of an overdense blob
(the cooling rate of the blob minus the cooling rate of the background
cooling-flow) is larger than the effective free-fall rate. The
effective free-fall rate is \be
\label{eq:eff_ff}
\frac{1}{t_{\rm ff, eff}} \sim \left ( \frac{\delta}{1+\delta} \right )^{1/2} \frac{1}{t_{\rm ff}},
\ee
where $\delta \equiv |\delta \rho/\rho| \sim |\delta T/T|$ is the overdensity of the blob relative to the ambient density. The effective cooling rate of an overdense blob is
\be
\label{eq:eff_cool}
\frac{1}{t_{\rm cool, eff}} = \frac{1}{t_{\rm cool, b}} - \frac{1}{t_{\rm cool, a}},
\ee
where $t_{\rm cool, b}$ is the cooling rate of the blob and $t_{\rm cool, a}$ is the cooling rate of the ambient medium. For small overdensities,  and in the isobaric limit, the effective cooling rate can be simplified to 
$$
\frac{1}{t_{\rm cool, eff}} \sim \delta  \frac{\partial}{\partial \ln T} \left (  \frac{-1}{t_{\rm cool}} \right )_p \sim \frac{5}{3} \frac{\delta}{t_{\rm TI}},
$$
where $t_{\rm TI}$ is the isobaric TI timescale given by
Eq. \ref{eq:tTI}. The effective free-fall rate for small overdensities
is $1/t_{\rm ff, eff} \sim \delta^{1/2}/t_{\rm ff}$, so that a cooling
flow can become multi-phase if \be
\label{eq:MP_crit}
t_{\rm TI} \lesssim \frac{5}{3} \delta^{1/2} t_{\rm ff} \ee~(again
assuming $\delta \lesssim 1$). Note also that a cooling-flow can become multi-phase only if the
cooling curve is thermally unstable in the isobaric regime; i.e., if
$d\ln \Lambda/d\ln T < 2$; when this condition is not met, the
slightly overdense blobs are heated relative to the background and
hence cannot become denser/cooler.

\begin{figure}
\centering
\includegraphics[scale=0.34]{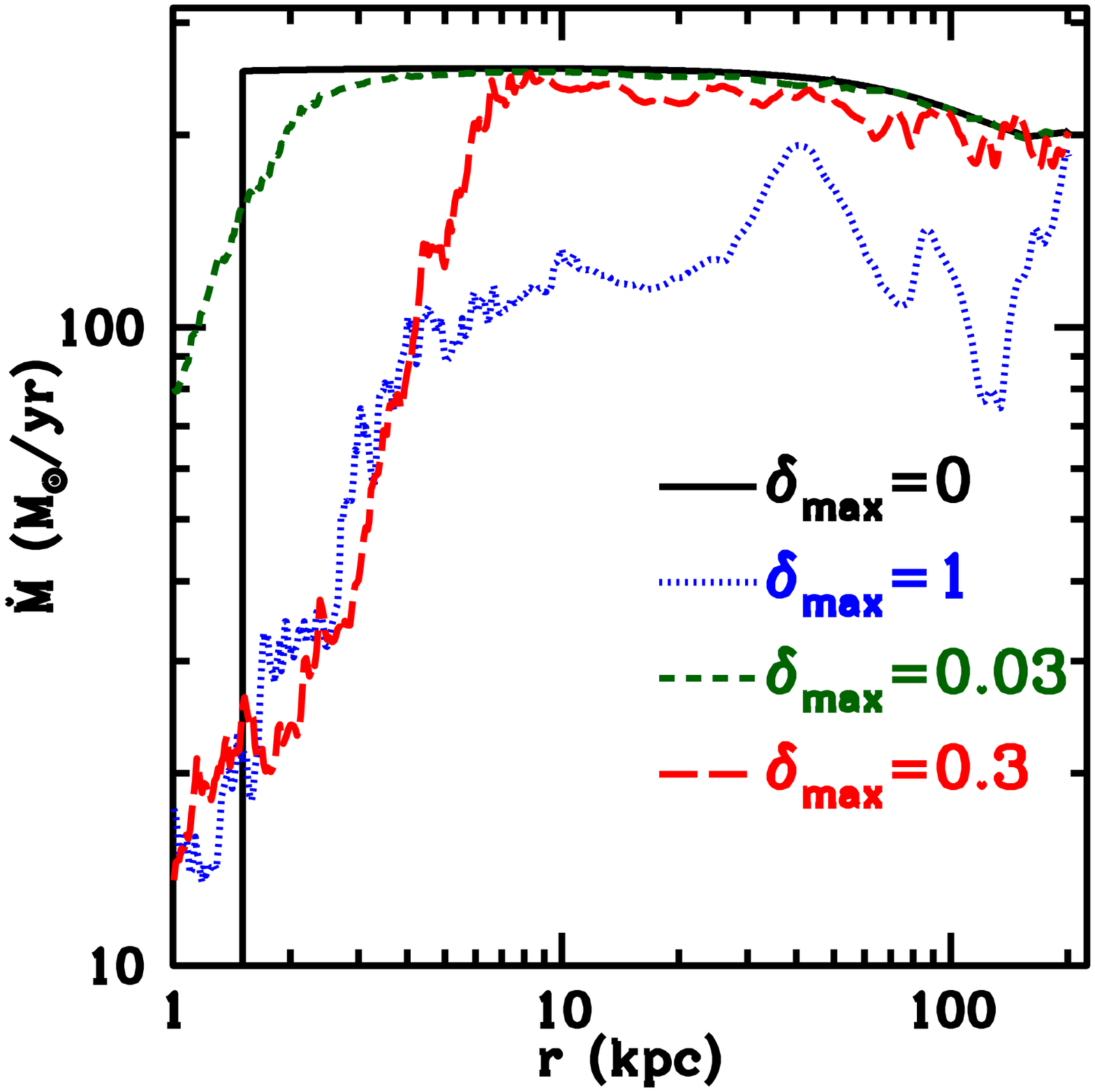}
\caption{Mass accretion rate in the hot phase~($T>0.1$~keV) as a function of radius averaged from 4 to 5~Gyr. 
The total accretion rate is relatively independent of both radius and the magnitude of the density perturbations (parameterized by $\delta_{\rm max}$; see Fig A1).  However, the mass accretion rate in the hot phase decreases at small radii by an amount that depends on the amplitude of the density perturbations:  larger density perturbations imply more gas cooling to low temperatures and thus a lower mass accretion rate in the hot phase. \label{fig:mdot_cf}}
  \end{figure}

The right panel of Figure~\ref{fig:cf_vs_r} shows the angle averaged
timescales in the quasi-steady state for a cooling-flow simulation
with density perturbations identical to the fiducial run~($\delta_{\rm max} \approx 0.3$); these timescales are calculated using the
properties of the hot plasma only~($T>0.1$~keV). The sound crossing
time for the hot plasma is $\lesssim$ the free-fall time at all radii,
implying that the hot plasma is in approximate hydrostatic equilibrium
at all radii. Within $r \lesssim 7$~kpc, however, the cooling time in
the hot plasma, which is  of order the infall time~(see
Eq. \ref{eq:en_cf}), is also of order the free-fall time.  While the
conservation laws of momentum and energy~(eqs. \ref{eq:mom_cf} \&
\ref{eq:en_cf}) are roughly valid when applied solely to the hot phase, the
mass conservation equation~(Eq. \ref{eq:mass_cf}) is
not. Specifically, mass is not conserved for the hot plasma within $r
\lesssim 7$~kpc because mass ``drops out'' via thermal instability to
the cold phase: the mass accretion rate in the hot plasma thus
decreases with a decreasing radius for $r \lesssim 7$~kpc. This is shown
explicitly in Figure~\ref{fig:mdot_cf} which shows the average mass accretion rate 
in the hot phase~($T>0.1$~keV) as a function of radius for the cooling-flow simulations
with different initial density perturbations.

The right panel of Figure~\ref{fig:cf_vs_r} also shows the cooling
timescale in the hot plasma in the quasi-steady state for the cooling
flow simulations initialized with density perturbations 3 times bigger
($\delta_{\rm max} = 1$) and 10 times smaller~($\delta_{\rm max} = 0.03$)
than our fiducial case.  The ratio of the cooling time to the
free-fall time at the radius where the cooling-flow becomes multi-phase
increases with larger density perturbations and the multi-phase cooling
flow sets in at larger radii~(see Eq. \ref{eq:MP_crit}).  

\begin{figure}
\centering
\includegraphics[scale=0.34]{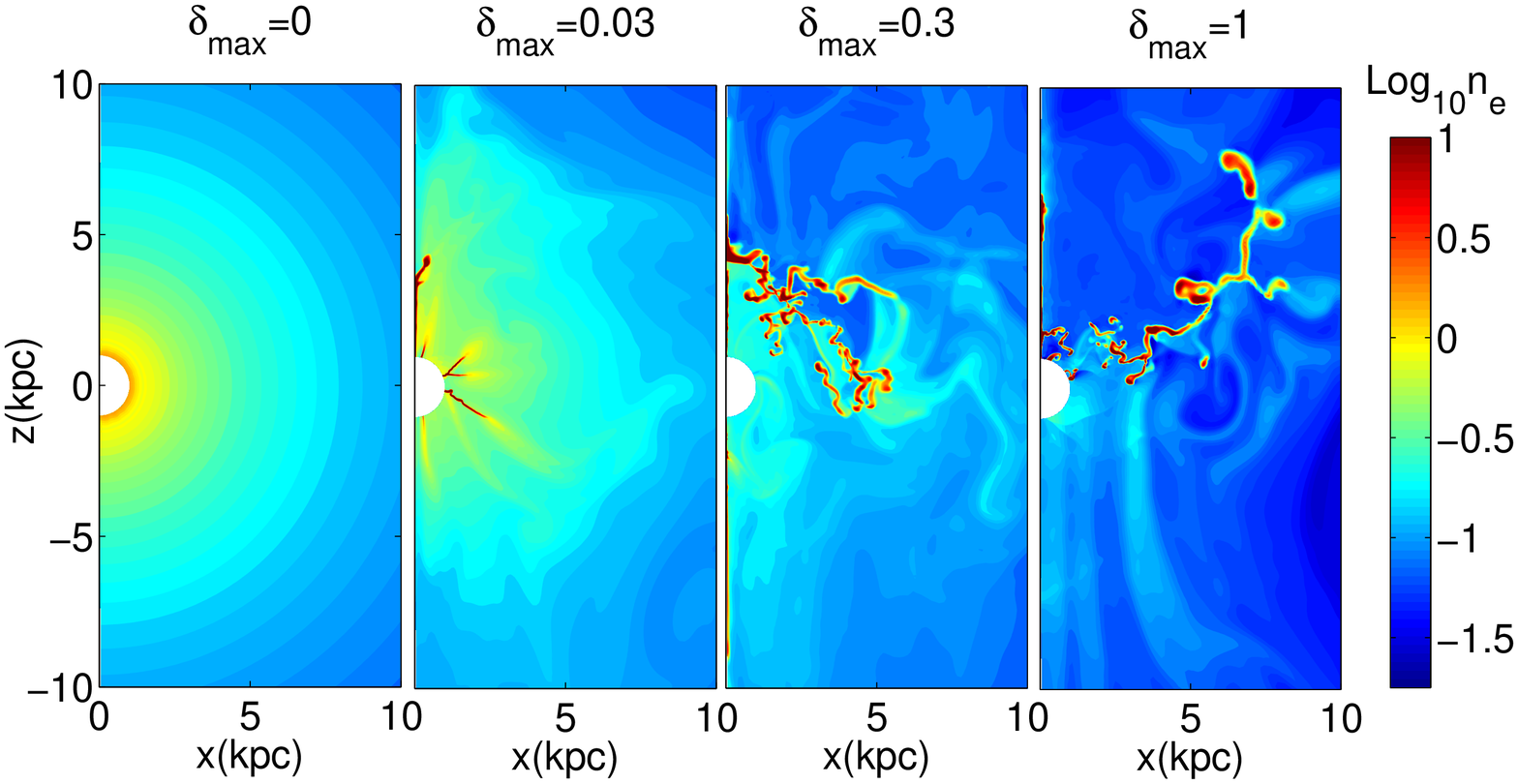}
\caption{Snapshots of the electron number density in the inner 10~kpc
  for the cooling-flow simulations with different fractional density
  perturbations $\delta \equiv |\delta \rho/\rho|$.  The simulations
  are initialized with a distribution of $\delta$ identical to our fiducial 
  run but rescaled such that $\delta
  < \delta_{\rm max}$.  This Figure demonstrates that a cooling-flow
  becomes multi-phase only when the cooling time is short and the
  density perturbations are sufficiently large~(see
  Eq. \ref{eq:MP_crit}).  The color scale for the electron density
  goes from $10^{-1.75}$ to 10~cm$^{-3}$ but the cold~($\sim 10^4$~K)
  gas in $\delta \neq 0$ simulations is actually much denser~($\sim 100-1000$~cm$^{-3}$).  \label{fig:2D_cf}} \end{figure}

Figure~\ref{fig:2D_cf} shows the electron number density in the inner
10~kpc for the quasi-steady state of cooling-flow simulations with
different initial density perturbations. Figure~\ref{fig:2D_cf} shows
the qualitative difference between a cooling-flow with and without
significant initial density perturbations: cooling-flows without significant
density perturbations do not show dense multi-phase gas~(consistent
with the analytic arguments of \citealt{bal89}); instead, the plasma just 
monolithically cools to low temperatures at radii where the
cooling time is less than the free-fall time~(see Fig.~\ref{fig:cf_vs_r}).  Figure~\ref{fig:2D_cf} also visually demonstrates
that the multi-phase structure that develops in cooling-flows with
large initial density perturbations depends strongly on the {\em
  magnitude} of the density perturbations: the models with larger
density perturbations develop more pronounced multi-phase structure out
to larger radii.  This is a critical difference between cooling-flow
models and models in which the plasma in hot halos is in approximate
global thermal equilibrium.  In the latter, the thermal instability
grows exponentially and the multi-phase structure that develops is
essentially independent of the initial density perturbations.  Thus
the development of multi-phase structure is far more generic and robust
if the virialized plasma is in approximate global thermal equilibrium.

Cooling-flow simulations are quite sensitive to the boundary condition at the outer radius because the free-fall time at the outer boundary is shorter than the Hubble time. A cosmologically realistic boundary condition should allow mass inflow at the outer boundary. Cooling-flow simulations which do not allow inflow, e.g., the isolated halo simulations of \citet{kau09}, evolve qualitatively differently from simulations with  inflow such as ours. Without mass supply at the outer boundary all the gas with cooling time 
 shorter than the age cools and accumulates at the center. Multiphase gas in such simulations occurs only for a short time comparable to the cooling time. In contrast, our cooling-flow simulations which allow inflow at the outer boundary reach a quasi-steady state with multiphase gas if density perturbations are large enough.

\label{lastpage}

\end{document}